\documentclass[runningheads]{llncs}

\makeatletter
\RequirePackage[bookmarks,unicode,colorlinks=true]{hyperref}%
   \def\@citecolor{blue}%
   \def\@urlcolor{blue}%
   \def\@linkcolor{blue}%

\def\orcidID#1{\smash{\href{http://orcid.org/#1}{\protect\raisebox{-1.25pt}{\protect\includegraphics{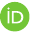}}}}}
\makeatother

\usepackage{booktabs}   
\usepackage{subcaption} 
\usepackage{xspace}
\usepackage{cite} 
\usepackage{comment} 
\usepackage{todonotes} 
\usepackage{listings} 
\usepackage{color} 
\usepackage{graphicx} 
\usepackage{subcaption}
\usepackage{paralist}
\usepackage{caption}
\usepackage{calc,booktabs,xparse,environ}
\usepackage{marvosym} 
\usepackage[inline]{enumitem}
\definecolor{red}{rgb}{0.6,0,0} 
\definecolor{blue}{rgb}{0,0,0.6}
\definecolor{green}{rgb}{0,0.8,0}
\definecolor{cyan}{rgb}{0.0,0.6,0.6}

\def\DZ#1{{\textcolor{blue}{\textbf{DZ:} #1}}}

\newcommand{\supplementary}{the supplementary material\xspace}

\newcommand{\tool}{\textsc{Paracosm}\xspace} 
\newcommand{\opendrive}{\textsc{OpenDRIVE}\xspace}
\newcommand{\openscenario}{\textsc{OpenSCENARIO}\xspace}

\DeclareCaptionFont{white}{\color{white}}
\DeclareCaptionFormat{listing}{\colorbox{blue}{\parbox{\textwidth}{\hspace{15pt}#1#2#3}}}

\def\set#1{{\{ #1 \}}}

\newcommand{\SOR}{&\mathrel{\makebox[\widthof{$\to$}]{$|$}}%
}

\ExplSyntaxOn
\NewEnviron{abstract_syntax}
 {
  \tl_replace_all:Nnn \BODY { \par } { }
  $\begin{aligned}
  \haskell_syntax:V \BODY
  \end{aligned}$
 }
\cs_new_protected:Nn \haskell_syntax:n
 {
  \seq_clear:N \l_haskell_syntax_output_seq
  \seq_set_split:Nnn \l_haskell_syntax_body_seq { \\ } { #1 }
  \seq_map_inline:Nn \l_haskell_syntax_body_seq
   {
    \seq_set_split:Nnn \l_haskell_syntax_row_seq { \SOR } { ##1 }
    \seq_put_right:Nx \l_haskell_syntax_output_seq
     {
      \seq_use:Nn \l_haskell_syntax_row_seq { \haskell_syntax_break: }
      \exp_not:N \\
     }
   }
  \seq_use:Nn \l_haskell_syntax_output_seq { \addlinespace }
 }
\cs_generate_variant:Nn \haskell_syntax:n { V }
\cs_new_protected:Nn \haskell_syntax_break: { \\ \SOR }
\ExplSyntaxOff

\def\calB{{\mathcal{B}}}

\begin{document}
\title{\tool: A Test Framework for Autonomous Driving Simulations}
\titlerunning{\tool: A Test Framework for Autonomous Driving Simulations}
%
\author{Rupak Majumdar\inst{1}\orcidID{0000-0003-2136-0542} \and
   Aman Mathur\inst{1}\orcidID{0000-0003-2405-0435} \Letter \and
   Marcus Pirron\inst{1}\orcidID{0000-0002-6501-728X} \and 
   Laura Stegner\inst{2}\orcidID{0000-0003-4496-0727} \and 
   Damien Zufferey\inst{1}\orcidID{0000-0002-3197-8736}}

\authorrunning{R. Majumdar et al.}


\institute{MPI-SWS, Kaiserslautern, Germany \email{\{rupak, mathur, mpirron, zufferey\}@mpi-sws.org} \and
University of Wisconsin-Madison, USA \email{stegner@wisc.edu}
}

\maketitle              
\begin{abstract}
Systematic testing of autonomous vehicles operating in complex real-world scenarios is a difficult and expensive problem.
We present \tool, a framework for writing systematic test scenarios for autonomous driving simulations.
\tool allows users to programmatically describe complex driving situations with specific features, e.g., 
road layouts and environmental conditions, as well as reactive temporal behaviors of other cars and pedestrians.
A systematic exploration of the state space, both for visual features and for reactive interactions with the environment is made possible. 
We define a notion of test coverage for parameter configurations based on combinatorial testing and low dispersion sequences.
Using fuzzing on parameter configurations, our automatic test generator can maximize coverage of various behaviors and find problematic cases. 
Through empirical evaluations, we demonstrate the capabilities of \tool in programmatically modeling parameterized test environments, and in finding problematic scenarios.

\end{abstract}


\section{Introduction}
\label{sec:Intro}

Building autonomous driving systems requires complex and intricate engineering effort.
At the same time, ensuring their reliability and safety is an extremely difficult task. 
There are serious public safety and trust concerns\footnote{https://www.weforum.org/agenda/2019/08/self-driving-vehicles-public-trust/}, aggravated by recent accidents involving autonomous cars\footnote{https://www.ntsb.gov/investigations/AccidentReports/Reports/HWY18MH010-prelim.pdf}.
Software in such vehicles combine well-defined tasks such as trajectory planning, steering, acceleration 
and braking, with underspecified tasks such as building a semantic model of the environment from raw sensor data and making decisions using this model.
Unfortunately, these underspecified tasks are critical to the safe operation of autonomous vehicles.
Therefore, testing in large varieties of realistic scenarios is the only way to build confidence in the correctness of the overall system.

Running real tests is a necessary, but slow and costly process.
It is difficult to reproduce corner cases due to infrastructure and safety issues; 
one can neither run over pedestrians to demonstrate a failing test case, nor wait for specific weather and road conditions.
Therefore, the automotive industry tests autonomous systems in virtual simulation environments \cite{Pomerleau88,url_torcs,url_UdacitySim,url_GazeboCarSim, airsim17,carla17}.
Simulation reduces the cost per test, and more importantly, gives precise control over all aspects of the environment, so as to test corner cases.

A major limitation of current tools is the lack of customizability: they either provide a GUI-based interface to design an environment piece-by-piece, or focus on bespoke pre-made environments.
This makes the setup of varied scenarios difficult and time consuming.
Though exploiting parametricity in simulation is useful and effective \cite{raja_ase, seshiaSemanticAdversarialDL, toyotaSim-basedAdversarial, jens_sbt_eval},
the cost of environment setup, and navigating large parameter spaces, is quite high \cite{jens_sbt_eval}.
Prior works have used bespoke environments with limited parametricity.
More recently, programmatic interfaces have been proposed \cite{seshiapldi19} to make such test procedures more systematic.
However, the simulated environments are largely still fixed, with no dynamic behavior.

\begin{figure}[t]
    \centering
  \includegraphics[width=0.85\textwidth]{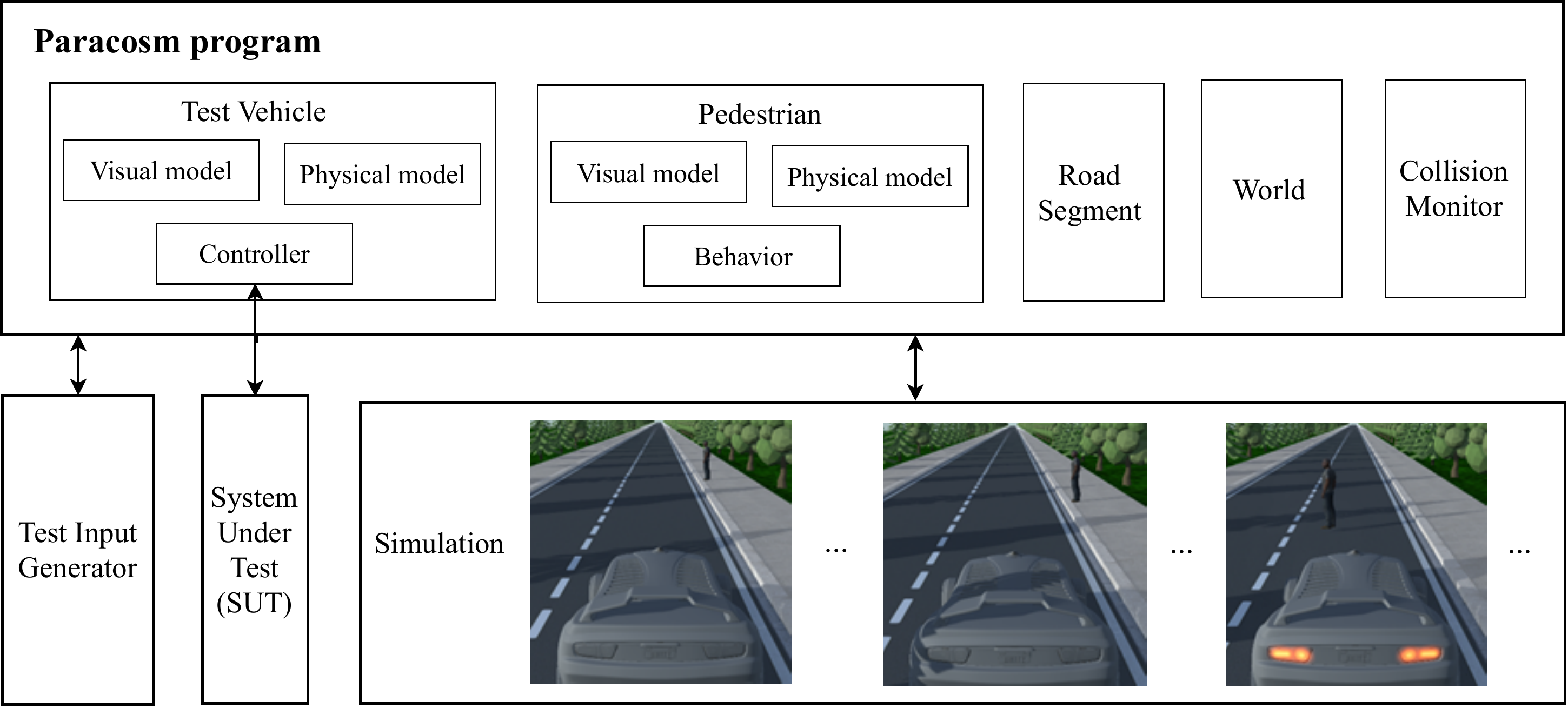}
  \caption{A \tool program consists of parameterized reactive components such as the test vehicle, the environment, road networks, other actors and their behaviors, and monitors.
The test input generation scheme guarantees good coverage over the parameter space.
The test scenario depicted here shows a test vehicle stopping for a jaywalking pedestrian.}
  \label{fig:ParacosmArchitecture2}
\vspace*{-1.8em}
\end{figure}

In this work, we present \tool, a programmatic interface that enables the design of \emph{parameterized environments} and \emph{test cases}.
Test parameters control the environment and the behaviors of the actors involved.
\tool supports various test input generation strategies, and we provide a notion of coverage for these.
Rather than computing coverage over intrinsic properties of the system under test (which is not yet understood for neural networks \cite{DBLP:conf/icse/LiM0C19}), our coverage criteria is over the space of test parameters.
Figure~\ref{fig:ParacosmArchitecture2} depicts the various parts of a \tool test.
A \tool program represents a family of tests, where each instantiation of the program's parameters is a concrete test case.



\tool is based on a synchronous reactive programming model \cite{DBLP:conf/pldi/WanH00,DBLP:conf/afp/HudakCNP02,DBLP:conf/popl/CaspiPHP87,RxNet}.
Components, such as road segments or cars, receive streams of inputs and produce streams of outputs over time.
In addition, components have graphical assets to describe their 
appearance for an underlying visual rendering engine and physical properties for an underlying physics simulator. 
For example, a vehicle in \tool not only has code that reads in sensor feeds and outputs steering angle or braking, but also has a textured mesh representing its shape, position and orientation in 3D space, and a physics model for its dynamical behavior.
A \tool configuration consists of a composition of several components.
Using a set of system-defined components (road segments, cars, pedestrians, etc.) combined using expressive operations from the 
underlying reactive programming model, users can set up complex temporally varying driving scenarios. 
For example, one can build an urban road network with intersections, pedestrians and vehicular traffic, and parameterize both,
environment conditions (lighting, fog), and behaviors (when a pedestrian crosses a street).  

Streams in the world description can be left ``open'' and, during testing, \tool automatically generates 
sequences of values for these streams.
We use a coverage strategy based on \emph{$k$-wise combinatorial coverage} \cite{Colbourn2004,Kuhn2010} 
for discrete variables and \emph{dispersion} for continuous variables.
Intuitively, $k$-wise coverage ensures that, for a programmer-specified parameter $k$, 
all possible combinations of values of any $k$ discrete parameters are covered by tests.
Low dispersion \cite{RoteTichy} ensures that there are no ``large empty holes'' left in the continuous parameter space.
\tool uses an automatic test generation strategy that offers high coverage based on random sampling over discrete 
parameters and \emph{deterministic} quasi-Monte Carlo methods for continuous parameters \cite{RoteTichy,Niederreiter}.

Like many of the projects referenced before, our implementation performs simulations inside a game engine.
However, \tool configurations can also be output to the \opendrive format \cite{opendrive} for use with other simulators, which is more in-line with the current industry standard.
We demonstrate through various case studies 
how \tool can be an effective testing framework for both qualitative properties (crash) and quantitative properties (distance maintained while following a car, or image misclassification).

Our main contributions are the following:
\begin{inparaenum}[(I)]
\item
    We present a programmable and expressive framework for programmatically modeling complex and parameterized scenarios to test autonomous driving systems. 
    Using \tool one can specify the environment's layout, behaviors of actors, and expose parameters to a systematic testing infrastructure.
\item
    We define a notion of test coverage based on combinatorial $k$-wise coverage in discrete space and low dispersion in continuous space.
    We show a test generation strategy based on fuzzing that theoretically guarantees good coverage.
\item
    We demonstrate empirically that our system is able to express complex scenarios and automatically 
    test autonomous driving agents and find incorrect behaviors or degraded performance.
\end{inparaenum}
%

\section{\tool through Examples}
\label{sec:LangThroughExamples}

We now provide a walkthrough of \tool through a testing example.
Suppose we have an autonomous vehicle to test. 
Its implementation is wrapped into a parameterized class:
\begin{lstlisting} [numbers=none, basicstyle=\small\ttfamily]
AutonomousVehicle(start, model, controller) { 
	void run(...) { ... } }
\end{lstlisting}
where the \lstinline{model} ranges over possible car models (appearance, physics),
and the \lstinline{controller} implements an autonomous controller.
The goal is to test this class in many different driving scenarios, including different road networks,
weather and light conditions, and other car and pedestrian traffic.
We show how \tool enables writing such tests as well as generate test inputs automatically.

A \emph{test configuration} consists of a composition of \emph{reactive objects}.
The following is an outline of a test configuration in \tool, in which the 
autonomous vehicle drives on a road with a pedestrian wanting to cross.
We have simplified the API syntax for the sake of clarity and omit the enclosing \lstinline{Test} class.
In the code segments, we use `\lstinline{:}' for named arguments.
\begin{lstlisting}[basicstyle=\small\ttfamily]
// Test parameters
light = VarInterval(0.2, 1.0) // value in [0.2, 1.0]
nlanes = VarEnum({2,4,6}) // value is 2, 4 or 6
// Description of environment
w = World(light:light, fog:0)
// Create a road segment
r = StraightRoadSegment(len:100, nlanes:nlanes)
// The autonomous vehicle controlled by the SUT
v = AutonomousVehicle(start:...,model:...,controller:...)
// Some other actor(s)
p = Pedestrian(start:.., model:..., ...)
// Monitor to check some property
c = CollisionMonitor(v) 
// Place elements in the world
run_test(env: {w, r, v, p}, test_params: {light, nlanes}, monitors: {c}, iterations: 100)
\end{lstlisting}
An instantiation of the reactive objects in the test configuration
gives a \emph{scene}---all the visual elements present in the simulated world.
A \emph{test case} provides concrete inputs to each ``open'' input stream in a scene.
A test case determines how the scene evolves over time: how the cars and pedestrians move and how environment conditions change.
We go through each part of the test configuration in detail below.

\paragraph{Reactive Objects.}
The core abstraction of \tool is a \emph{reactive object}.
Reactive objects capture geometric 
and graphical features of a physical object, 
as well as their behavior over time.
The behavioral interface for each reactive object has a set of \emph{input} streams and a set of \emph{output} streams.
The evolution of the world is computed in steps of fixed duration which corresponds to events in a predefined \lstinline{tick} stream.
For streams that correspond to physical quantities updated by the physics simulator, such as position and speeds of cars, etc., 
appropriate events are generated by the underlying physics simulator.

Input streams provide input values from the environment over time; output streams represent output values computed by the object.
The object's constructor sets up the internal state of the object.
An object is updated by event triggered computations.
\tool provides a set of assets as base classes.
Autonomous driving systems naturally fit reactive programming models.
They consume sensor input streams and produce actuator streams for the vehicle model.
We differentiate between static \emph{environment} reactive objects (subclassing \lstinline{Geometric}) 
and dynamic \emph{actor} reactive objects (subclassing \lstinline{Physical}).
Environment reactive objects represent ``static'' components of the world, such 
as road segments, intersections, buildings or trees, and a special component called the \emph{world}.
Actor reactive objects represent components with ``dynamic'' behavior: vehicles or pedestrians.
The world object is used to model features of the world such as lighting or weather conditions.
Reactive objects can be \emph{composed} to generate complex assemblies from simple objects.
The composition process can be used to connect static components structurally--such as two road segments connecting at an intersection.
Composition also connects the behavior of an object to another by binding output streams to input streams.
At run time, the values on that input stream of the second object are obtained from the output values of the first. 
Composition must respect geometric properties---the runtime system ensures that a composition maintains invariants such as no intersection of geometric components.
We now describe the main features in \tool, centered around the test configuration above.

\begin{figure}[t]
	\centering
	\includegraphics[width=0.5\textwidth]{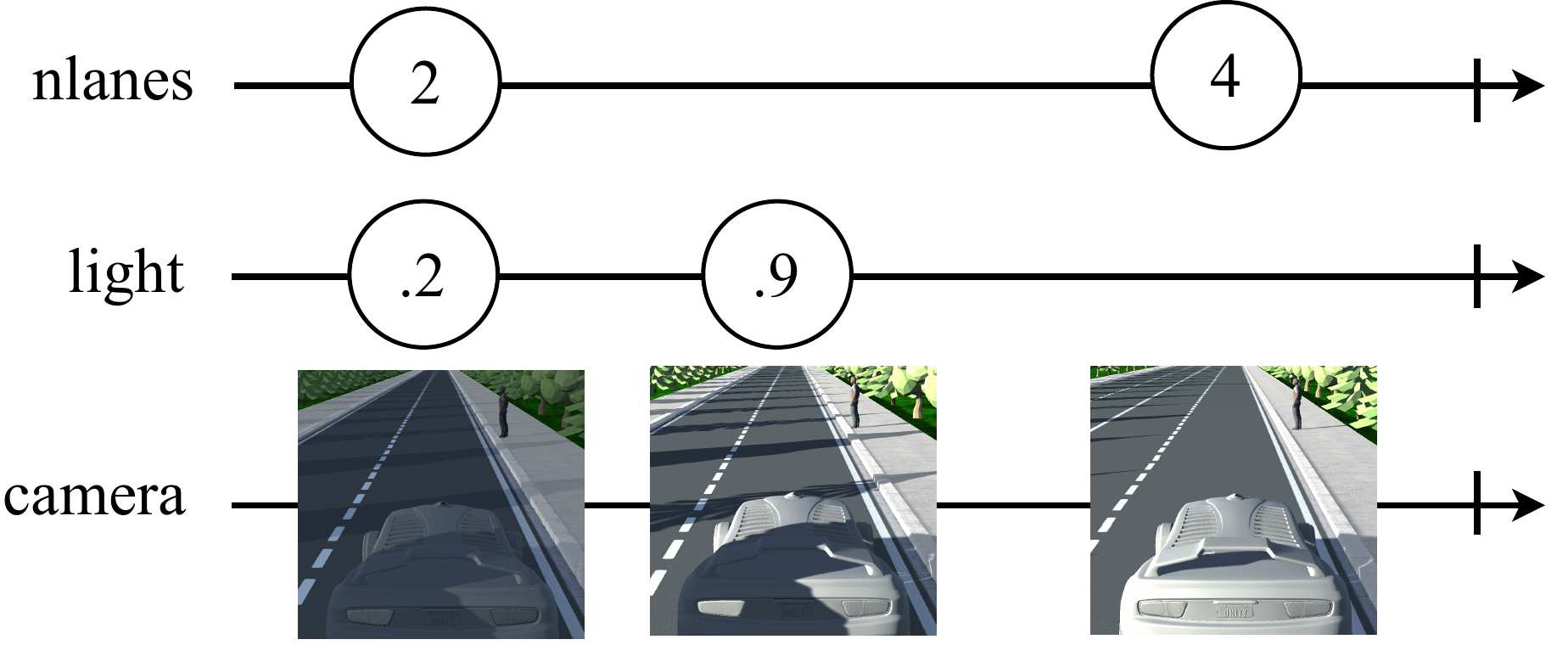}
  	\caption{Reactive streams represented by a marble diagram. A change in the value of test parameters \lstinline{nlanes} or \lstinline{light} changes the environment, and triggers a change in the corresponding sensor (output) stream \lstinline{camera}.}
  \label{fig:ReactiveVars}
\vspace*{-1em}
\end{figure}

\paragraph{Test Parameters.}
Using test variables, we can have general, but constrained streams of values passed into objects \cite{Millstein}.
Our automatic test generator can then pick values for these variables, thereby leading to different test cases (see Figure~\ref{fig:ReactiveVars}).
There are two types of parameters: continuous (\lstinline{VarInterval}) and discrete (\lstinline{VarEnum}).
In the example presented, \lstinline{light} (light intensity) is a continuous test parameter and \lstinline{nlanes} (number of lanes) is discrete.

\paragraph{World.}
The \lstinline{World} is a pre-defined reactive object in \tool
with a visual representation responsible for atmospheric conditions like the light intensity, direction 
and color, fog density, etc.
The code segment
\begin{lstlisting}[numbers=none, basicstyle=\small\ttfamily]
w = World(light:light, fog:0)
\end{lstlisting}
parameterizes the world using a test variable for light and sets the fog density to a constant (0).

\paragraph{Road Segments.}
In our example, \lstinline{StraightRoadSegment} was parameterized with the number of lanes.
In general, \tool provides the ability to build complex road networks by connecting primitives of individual road segments and intersections.
(A detailed example is presented in the Appendices.)

It may seem surprising that we model static scene components such as roads as reactive objects. 
This serves two purposes.
First, we can treat the number of lanes in a road segment as a constant input stream that 
is set by the test case, allowing parameterized test cases.
Second, certain features of static objects can also change over time.
For example, the coefficient of friction on a road segment may depend on the weather condition, which can be a function of time.

\paragraph{Autonomous Vehicles \& System Under Test (SUT).}
\lstinline{AutonomousVehicle}, as well as other actors, 
extends the \lstinline{Physical} class (which in turn subclasses \lstinline{Geometric}).
This means that these objects have a visual as well as a physical model.
The visual model is essentially a textured 3D mesh.
The physical model contains properties such as mass, moments of inertia of separate bodies in the vehicle, joints, etc.
This is used by the physics simulator to compute the vehicle's motion in response to external forces and control input.
In the following code segment, we instantiate and place our test vehicle on the road:
\begin{lstlisting}[numbers=none, basicstyle=\small\ttfamily]
v = AutonomousVehicle(start:r.onLane(1, 0.1), model:CarAsset(...), controller:MyController(...))
\end{lstlisting}
The \lstinline{start} parameter ``places'' the vehicle in the world (in relative coordinates).
The \lstinline{model} parameter provides the implementation of the geometric and physical model of the vehicle.
The \lstinline{controller} parameter implements the autonomous controller under test.
The internals of the controller implementation are not important; what is important is its interface (sensor inputs
and the actuator outputs).
These determine the input and output streams that are passed to the controller during simulation.
For example, a typical controller can take sensor streams such as image streams from a camera as input
and produce throttle and steering angles as outputs. 
The \tool framework ``wires'' these streams appropriately.
For example, the rendering engine determines the camera images based on the geometry of the scene and the position
of the camera and the controller outputs are fed to the physics engine to determine the updated scene.
Though simpler systems like \textsc{openpilot} \cite{openpilot} use only a dashboard-mounted camera, autonomous vehicles can, in general, mix cameras at various mount points, LiDARs, radars, and GPS.
\tool can emulate many common types of sensors which produce streams of data.
%
It is also possible to integrate new sensors, which are not supported out-of-the-box, by implementing them using the game engine's API.

\paragraph{Other Actors.}
A test often involves many actors such as pedestrians, and other (non-test) vehicles.
Apart from the standard geometric (optionally physical) properties, these can also have some pre-programmed behavior.
Behaviors can either be only dependent on the starting position (say, a car driving straight on the same lane), or be dynamic and reactive,
depending on test parameters and behaviors of other actors.
In general, the reactive nature of objects enables complex scenarios to be built.
For example, here, we specify a simple behavior of a pedestrian crossing a road.
The pedestrian starts crossing the road when a car is a certain distance away.
In the code segments below, we use `\lstinline{_}' as shorthand for a lamdba expression, i.e., ``\lstinline{f(_)}'' is the same as ``\lstinline{x => f(x)}''.

\begin{lstlisting} [numbers=none, basicstyle=\small\ttfamily]
Pedestrian(value start, value target, carPos, value dist, value speed) extends Geometric {
  ... // Initialization
  // Generate an event when the car gets close
  trigger = carPos.Filter( abs(_ - start) < dist )
  // target location reached
  done = pos.Filter( _ == target )
  // Walk to the target after trigger fires
  tick.SkipUntil(trigger).TakeUntil(done).foreach( ... /* walk with given speed */ )
}
\end{lstlisting}

\paragraph{Monitors and Test Oracles.}
\tool provides an API to provide qualitative and quantitative temporal specifications.
For instance, in the following example, we check that there is no collision and ensure that the collision was not trivially avoided because our vehicle did not move at all.

\begin{lstlisting} [numbers=none, basicstyle=\small\ttfamily]
// no collision
CollisionMonitor(AutonomousVehicle v) extends Monitor {
  assert(v.collider.IsEmpty()) }
// cannot trivially pass the test by staying put
DistanceMonitor(AutonomousVehicle v, value minD) extends Monitor {
  pOld = v.pos.Take(1).Concat(v.pos)
  D = v.pos.Zip(pOld).Map( abs(_ - _) ).Sum()
  assert(D >= minD)
}
\end{lstlisting}
The ability to write monitors which read streams of system-generated events provides an expressive framework to write temporal properties, something that has been identified as a major limitation of prior tools \cite{jens_sbt_eval}.
Monitors for metric and signal temporal logic specifications can be encoded in the usual way \cite{DBLP:conf/rv/HoOW14,DBLP:journals/fmsd/DeshmukhDGJJS17}.

\section{Systematic Testing of \tool Worlds}
\label{sec:TestInterface}

\subsection{Test Inputs and Coverage}

Worlds in \tool directly describe a parameterized family of tests.
The testing framework allows users to specify various strategies to generate 
input streams for both, static, and dynamic reactive objects in the world.

\paragraph{Test Cases.}
A \emph{test} of \emph{duration} $T$ executes a configuration of reactive objects by 
providing inputs to every open input stream in the configuration for $T$ ticks.
The inputs for each stream must satisfy \lstinline{const} parameters and respect the range constraints from \lstinline{VarInterval} and \lstinline{VarEnum}.
The runtime system manages the scheduling of inputs and pushing input streams to 
the reactive objects.
Let $\mathsf{In}$ denote the set of all input streams, and 
$\mathsf{In} = \mathsf{In}_D \cup \mathsf{In}_C$ denote the partition of $\mathsf{In}$
into \emph{discrete} streams and \emph{continuous} streams respectively.
Discrete streams take their value over a finite, discrete range; for example, the color
of a car, the number of lanes on a road segment, or the position of the next pedestrian (left/right)
are discrete streams.
Continuous streams take their values in a continuous (bounded) interval.
For example, the fog density or the speed of a vehicle
are examples of continuous streams.

\paragraph{Coverage.}
In the setting of autonomous vehicle testing, one often wants to explore the state space
of a parameterized world to check ``how well'' an autonomous vehicle works under various situations,
both qualitatively and quantitatively.
Thus, we now introduce a notion of coverage.
Instead of structural coverage criteria such as line or branch coverage, our goal is to cover the parameter 
space.
In the following, for simplicity of notation, we assume that all discrete streams take values from $\set{0,1}$,
and all continuous streams take values in the real interval $[0,1]$.
Any input stream over bounded intervals---discrete or continuous---can be encoded into such streams.
For discrete streams, there are finitely many tests, since each co-ordinate is Boolean and there is a fixed number of co-ordinates. 
One can define the coverage as the fraction of the number of vectors tested to the total number of vectors.
Unfortunately, the total number of vectors is very high: if each stream is constant, then there are already $2^{n}$ tests for $n$ streams.
Instead, we consider the notion of \emph{$k$-wise testing} from combinatorial testing \cite{Kuhn2010}.
In $k$-wise testing, we fix a parameter $k$, and ask that every interaction between every $k$ elements is tested.
Let us be more precise.
Suppose that a test vector has $N$ co-ordinates, where each co-ordinate can get the value $0$ or $1$.
A set of tests $A$ is a \emph{$k$-wise covering family} 
if for every subset $\set{i_1, i_2,\ldots, i_k} \subseteq \set{1,\ldots, N}$ of co-ordinates
and every vector $v\in \set{0,1}^k$, there is a test $t\in A$ whose restriction to the $i_1,\ldots, i_k$ is precisely $v$.

For continuous streams, the situation is more complex: since any continuous interval has infinitely many points,
each corresponding to a different test case, we cannot directly define coverage as a ratio (the denominator will be infinite).
Instead, we define coverage using the notion of \emph{dispersion} \cite{RoteTichy,Niederreiter}.
Intuitively, dispersion measures the largest empty space left by a set of tests.
We assume a (continuous) test is a vector in $[0,1]^N$: each entry is picked from the interval $[0,1]$ and there are $N$ co-ordinates.
Dispersion over $[0,1]^N$ can be defined relative to sets of neighborhoods, such as $N$-dimensional balls or axis-parallel rectangles.
Let us define $\calB$ to be the family of $N$-dimensional axis-parallel rectangles in $[0,1]^N$, our results also hold for other
notions of neighborhoods such as balls or ellipsoids.
For a neighborhood $B\in\calB$, let $\mathit{vol}(B)$ denote the volume of $B$.
Given a set $A \subseteq [0,1]^N$ of tests, we define the \emph{dispersion}
as the largest volume neighborhood in $\calB$ without any test: 
\[
\mathsf{dispersion}(A) = \sup\set{\mathrm{vol}(B) \mid B\in\calB \mbox{ and } A \cap B = \emptyset}
\]
A lower dispersion means better coverage.

Let us summarize.
Suppose that a test vector consists of $N_D$ discrete co-ordinates and $N_C$ continuous co-ordinates;
that is, a test is a vector $(t_D, t_C)$ in $\set{0,1}^{N_D} \times [0, 1]^{N_C}$.
We say a set of tests $A$ is \emph{$(k, \varepsilon)$-covering} if 
\begin{enumerate}
\item
for each set of $k$ co-ordinates $\set{i_1,\ldots, i_k} \subseteq \set{1,\ldots, N_D}$ and each vector $v\in \set{0,1}^k$,
there is a test $(t_D, t_C) \in \set{0,1}^{N_D} \times [0,1]^{N_C}$ such that 
the restriction of $t_D$ to the co-ordinates $i_1,\ldots, i_k$ is $v$; and

\item 
for each $(t_D, t_C)\in A$, the set $\set{t_C \mid (t_D, t_C)\in A}$ has dispersion at most $\epsilon$.
\end{enumerate}

\subsection{Test Generation}
\label{sec:test_generation} 
%
The goal of our default test generator is to maximize $(k, \epsilon)$ for programmer-specified number of test iterations or \lstinline{ticks}.

\paragraph{$k$-Wise Covering Family.}
One can use explicit construction results from combinatorial testing to generate $k$-wise covering families \cite{Colbourn2004}.
However, a simple way to generate such families with high probability is random testing.
The proof is by the probabilistic method \cite{Alon} (see also \cite{MN18}).
Let $A$ be a set of $2^k(k \log N - \log \delta)$ uniformly randomly generated $\set{0,1}^N$ vectors.
Then $A$ is a $k$-wise covering family with probability at least $1-\delta$.


\paragraph{Low Dispersion Sequences.}
It is tempting to think that uniformly generating vectors from $[0,1]^N$ would similarly give low dispersion sequences.
Indeed, as the number of tests goes to infinity, the set of randomly generated tests has dispersion 
$0$ almost surely.
However, when we fix the number of tests,
it is well known that uniform random sampling can lead to high dispersion \cite{Niederreiter,RoteTichy}; in fact, one
can show that the dispersion of $n$ uniformly randomly generated tests grows asymptotically as $O((\log \log n/n)^{\frac{1}{2}})$
almost surely.
Our test generation strategy is based on \emph{deterministic quasi-Monte Carlo sequences}, which have 
much better dispersion properties, 
asymptotically of the order of $O(1/n)$, than the dispersion behavior of uniformly random tests.
There are many different algorithms for generating quasi-Monte Carlo sequences deterministically (see, e.g., \cite{Niederreiter,RoteTichy}).
We use \emph{Halton sequences}.
%
For a given $\epsilon$, we need to generate $O(\frac{1}{\epsilon})$ inputs via Halton sampling.
In Section~\ref{sec:CaseStudies}, we compare uniform random and Halton sampling.

\paragraph{Cost Functions and Local Search.}
In many situations, testers want to optimize parameter values for a specific function.
A simple example of this is finding higher-speed collisions, which 
intuitively, can be found in the vicinity of test parameters that already result in high-speed collisions.
Another, slightly different case is (greybox) fuzzing \cite{url_afl,Rawat2017VUzzerAE}, for example, finding new collisions using small mutations on parameter values that result in the vehicle narrowly avoiding a collision.
Our test generator supports such \emph{quantitative} objectives and \emph{local search}. 
A quantitative monitor evaluates a cost function on a run of a test case.
Our test generation tool generates an initial, randomly chosen, set of test inputs.
Then, it considers the scores returned by the Monitor on these samples, 
and performs a local search on samples with the highest/lowest scores to find local optima of the cost function.

\section{Implementation and Tests}
\label{sec:ImplementationAndTests}

\subsection{Runtime System and Implementation}
\label{sec:RuntimeSystem}

\tool uses the Unity game engine \cite{url_unity3D} to render visuals, do runtime checks and simulate physics (via PhysX \cite{physx}).
Reactive objects are built on top of UniRx \cite{unirx}, an implementation of the popular Reactive Extensions framework \cite{url_rx}.
The game engine manages geometric transformations of 3D objects and offers easy to use abstractions for generating realistic simulations.
Encoding behaviors and monitors, management of 3D geometry and dynamic checks are implemented using the game engine interface.

A simulation in \tool proceeds as follows.
A test configuration is specified as a subclass of the \lstinline{EnvironmentProgramBaseClass}.
Tests are run by invoking the \lstinline{run_test} method, which receives as input the reactive objects
that should be instantiated in the world as well as additional parameters relating to the test. 
The \lstinline{run_test} method runs the tests by first initializing and placing the reactive objects in the
scene using their 3D mesh (if they have one) and then
invoking a reactive engine to start the simulation. 
The system under test is run in a separate process and connects to the simulation.
The simulation then proceeds until the simulation completion criteria is met (a time-out or some monitor event).

\paragraph{Output to Standardized Testing Formats.}
There have been recent efforts to create standardized descriptions of tests in the automotive
industry.
The most relevant formats are \opendrive \cite{opendrive} and \openscenario (only recently finalized) \cite{openscenario}.
\opendrive describes road structures, and \openscenario describes actors and their behavior.
\tool currently supports outputs to \opendrive.
Due to the static nature of the specification format, a different file is generated for each test iteration/configuration.

\subsection{Evaluation}
\label{sec:CaseStudies}

We evaluate \tool with respect to the following research questions (\textbf{RQ}s):\\
\textbf{RQ 1}: Does \tool's programmatic interface enable the easy design of test environments and worlds?\\
\textbf{RQ 2}: Do the test input generation strategies discussed in Section~\ref{sec:TestInterface} effectively explore the parameter space?\\
\textbf{RQ 3}: Can \tool help uncover poor performance or bad behavior of the SUT in common autonomous driving tasks?

\paragraph{Methodology.}
To answer \textbf{RQ 1}, we develop three independent environments rich with visual features and other actors, and use the variety generated with just a few lines of code as a proxy for ease of design.
To answer \textbf{RQ 2}, we use coverage maximizing strategies for test inputs to all the three environments/case studies.
We also use and evaluate cost functions and local search based methods.
To answer \textbf{RQ 3}, we test various neural network based systems and demonstrate how \tool can help uncover problematic scenarios.
A summary of the case studies presented here is available in Table~\ref{tab:case_study_summary}.
In the Appendices, we present more case studies, specifically experiments on many pre-trained neural networks, busy urban environments and studies exploiting specific testing features of \tool.

\begin{table}[t]
\caption{An overview of our case studies.
Note that even though the Adaptive Cruise Control study has 2 discrete parameters, we calculate k-wise coverage for 3 as the 2 parameters require 3 bits for representation.}
\label{tab:case_study_summary}
\begin{tabular}{p{1.2cm} p{3.15cm} p{3.45cm} p{3.5cm}} 
\toprule
 & Road segmentation & Jaywalking pedestrian & Adaptive Cruise Control\\
\toprule
SUT & VGGNet CNN \cite{Simonyan14verydeep} & NVIDIA CNN \cite{nvidiaEndToEnd} & NVIDIA CNN \cite{nvidiaEndToEnd}\\

Training & 191 images& 403 image \& car control samples & 1034 image \& car control samples\\

Test params & 3 discrete & 2 continuous & 3 continuous, 2 discrete \\

Test iters & 100 & 100, 15\emph{s} timeout & 100, 15\emph{s} timeout\\

Monitor & Ground truth & Scored Collision & Collision \& Distance \\

Coverage & $k = 3$ with probability $\sim 1$ & $\epsilon = 0.041$ & $\epsilon = 0.043$, $k = 3$ with probability $\sim 1$ \\
\bottomrule
\end{tabular}
\end{table}

\subsection{Case Studies}
\label{sec:road_segmentation}

\paragraph{Road segmentation}
\begin{figure}[t]
\centering
    \begin{subfigure}{0.47\textwidth}
        \centering
        \includegraphics[width=0.99\textwidth]{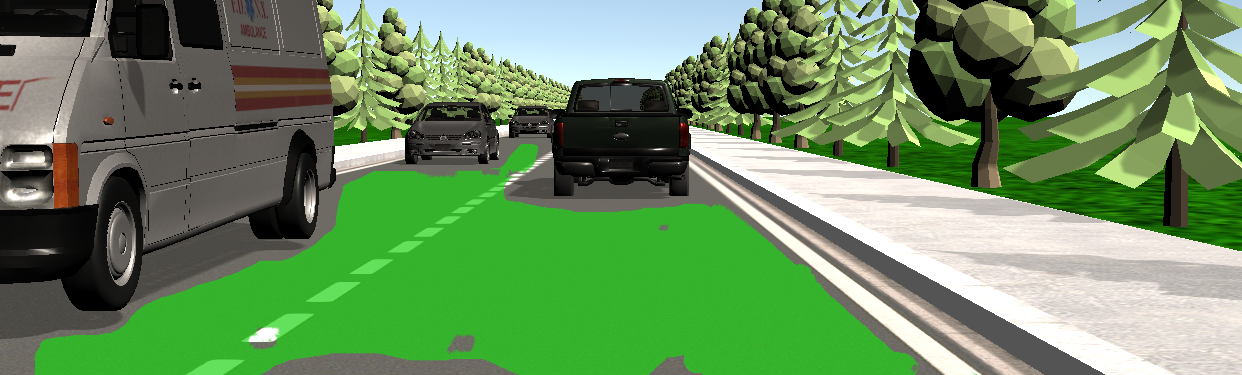}
        \label{fig:road_seg_training}
        \caption{A good test with all parameter values same as the training set (true positive: 89\%, false positive: 0\%).}
    \end{subfigure}
    \hfill
    \begin{subfigure}{0.47\textwidth}
        \centering
        \includegraphics[width=0.99\textwidth]{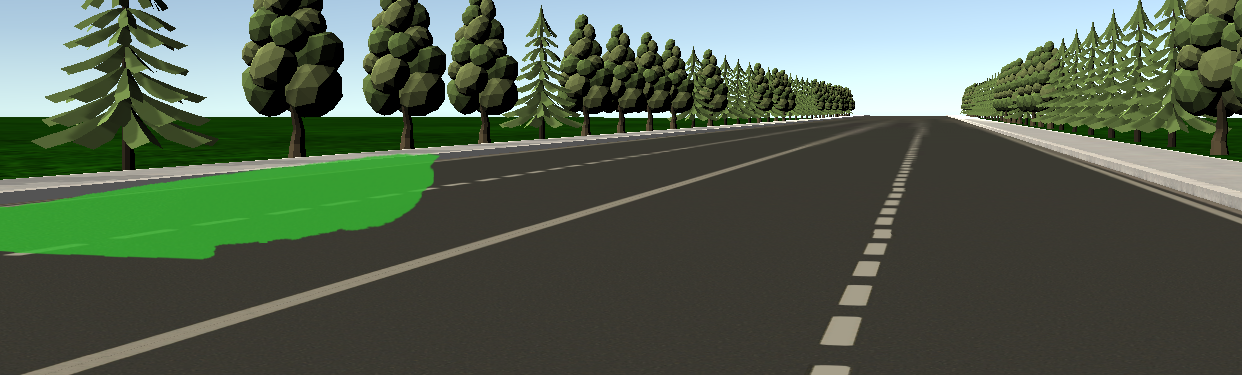}
        \label{fig:road_seg_test}
        \caption{A bad test with all parameter values different from the training set (true positive: 9\%, false positive: 1\%).}
    \end{subfigure}
\caption{Example results from the road segmentation case study. Pixels with a green mask are segmented by the SUT as a road.}
\label{fig:road_seg}
\end{figure}

\begin{table}[t]
\centering
        \caption{Summary of results of the road segmentation case study. 
    Each combination of parameter values is presented separately, with the parameter values used for training in bold.
    We report the SUT's average true positive rate (\% of pixels corresponding to the road that are correctly classified) and false positive rate (\% of pixels that are not road, but incorrectly classified as road).}
    \label{tab:kittiseg_summary}
    \begin{tabular}{ c c c c c c }
    \toprule
        \# lanes           &    \# cars &   Lighting   &   \# test iters             &         True positive (\%)  &       False positive (\%)  \\ \midrule
                   \textbf{2}               &    \textbf{5}  &   \textbf{Noon}      &       12                 &              70\%                      &                 5.1\%                 \\
        2               &    5  &   Evening   &       14                 &              53.4\%                       &                22.4\% \\
        2               &    0   &   Evening   &       12                 &              51.4\%                       &                18.9\%                 \\
        2               &    0   &   Noon      &       12                 &              71.3\%                       &                 6\%                 \\
        4               &    5  &   Evening   &       10                 &              60.4\%                       &                 7.1\%                 \\
        4               &    5  &   Noon      &       16                 & 	          68.5\%                       &                20.2\%                 \\
        4               &    0   &   Evening   &       13                 &              51.5\%                       &                 7.1\%                 \\ 
        4               &    0   &   Noon      &       11                 &              83.3\%                       &                21\%                 \\
        \bottomrule

    \end{tabular}
\end{table}
Using \tool's programmatic interface, we design a long road segment with several vehicles.
The vehicular behavior is to drive on their respective lanes with a fixed maximum velocity.
The test parameters are the number of lanes ($\{2,\:  4\}$), number of cars in the environment ($\{0,\:  5\}$) and light conditions ($\{Noon,\:  Evening\}$).
Noon lighting is  much brighter than the evening.
The direction of lighting is also the opposite.
We test a deep CNN called VGGNet \cite{Simonyan14verydeep}, that is known to perform well on several image segmentation benchmarks.
The task is road segmentation, i.e., given a camera image, identifying which pixels correspond to the road.
The network is trained on 191 dashcam images captured in the test environment with fixed parameters ($2$ lanes, $5$ cars, and $Noon$ lighting), recorded at the rate of one image every $1/10^{th}$ second, while manually driving the vehicle around (using a keyboard).
We test on 100 images generated using \tool's default test generation strategy (uniform random sampling for discrete parameters).
Table \ref{tab:kittiseg_summary} summarizes the test results.
Tests with parameter values far away from the training set are observed to not perform so well.
As depicted in Figure~\ref{fig:road_seg}, this happens because varying test parameters can drastically change the scene.

\paragraph{Jaywalking pedestrian.}
\begin{table}[t]
\centering
\footnotesize
\caption{Results for the jaywalking pedestrian case study.}
\label{tbl:ped_crossing}
\begin{tabular}{lrrrr}
\toprule
Testing strategy & Dispersion  ( $\epsilon$)  & \% fail  & Max. collision    \\ \midrule
Random     & 0.092   & $7\%$     & 10.5 m/s       \\ 
Halton      & 0.041  & $10\%$     & 11.3 m/s       \\ 
Random+opt/collision      & 0.109  & $13\%$    & 11.1 m/s      \\
Halton+opt/collision      & 0.043  & $20\%$    & 11.9 m/s      \\  
Random+opt/almost failing & 0.126 & $13\%$ & 10.5 m/s\\
Halton+opt/almost failing & 0.043 & $13\%$ & 11.4 m/s\\
\bottomrule
\end{tabular}
\end{table}
We now test over the environment presented in Section~\ref{sec:LangThroughExamples}.
The environment consists of a straight road segment and a pedestrian.
The pedestrian's behavior is to cross the road at a specific walking speed when the autonomous vehicle is a specific distance away.
The walking speed of the pedestrian and the distance of the autonomous vehicle when the pedestrian starts crossing the road are test parameters.
The SUT is a CNN based on NVIDIA's behavioral cloning framework \cite{nvidiaEndToEnd}.
It takes camera images as input, and produces the relevant steering angle or throttle control as output.
The SUT is trained on 403 samples obtained by driving the vehicle manually and recording the camera and corresponding control data.
The training environment has pedestrians crossing the road at various time delays, but always at a fixed walking speed (1 m/s).
In order to evaluate \textbf{RQ 2} completely, we evaluate the default coverage maximizing sampling approach, as well as explore two quantitative objectives: first, maximizing the collision speed, and second, finding new failing cases around samples that \emph{almost} fail.
For the default approach, the  \lstinline{CollisionMonitor} as presented in Section~\ref{sec:LangThroughExamples} is used. 
For the first quantitative objective, this \lstinline{CollisionMonitor}'s code is prepended with the following calculation:
\begin{lstlisting}[numbers=none, basicstyle=\small\ttfamily]
// Score is speed of car at time of collision 
coll_speed = v.speed.CombineLatest(v.collider, (s,c) => s).First()
\end{lstlisting}
The score \lstinline{coll_speed} is used by the test generator for optimization.
For the second quantitative objective, the \lstinline{CollisionMonitor} is modified to give high scores to tests where the distance between the autonomous vehicle and pedestrian is very small:
\begin{lstlisting} [numbers=none, basicstyle=\small\ttfamily]
CollisionMonitor(AutonomousVehicle v, Pedestrian p) extends Monitor {
	minDist = v.pos.Zip(p.pos).Map(1/abs(_-_)).Min()
	coll_score = v.collider.Map(0)
	// Score is either 0 (collision) or 1/minDist
	score = coll_score.DefaultIfEmpty(minDist)
	assert(v.collider.IsEmpty())
}
\end{lstlisting}

We evaluate the following test input generation strategies:
\begin{enumerate*}[label=(\roman*)]
\item Random sampling 
\item Halton sampling,
\item Random or Halton sampling with local search for the two quantitative objectives.
\end{enumerate*}
We run 100 iterations of each strategy with a 15 second timeout.
For random or Halton sampling, we sample 100 times.
For the quantitative objectives, we first generate 85 random or Halton samples, then choose the top 5 scores, and finally run 3 simulated annealing iterations on each of these 5 configurations.
Table~\ref{tbl:ped_crossing} presents results from the various test input generation strategies.
Clearly, Halton sampling offers the lowest dispersion (highest coverage) over the parameter space. 
This can also be visually confirmed from the plot of test parameters (Figure~\ref{fig:Pedestrian_halton_100}).
There are no big gaps in the parameter space.
Moreover, we find that test strategies optimizing for the first objective are successful in finding more collisions with higher speeds.
As these techniques perform simulated annealing repetitions on top of already failing tests, they also find more failing tests overall.
Finally, test strategies using the second objective are also successful in finding more (newer) failure cases than simple Random or Halton sampling.

\begin{figure}[t]
\centering
    \begin{subfigure}{0.31\textwidth}
    \centering
        \includegraphics[width=0.99\textwidth]{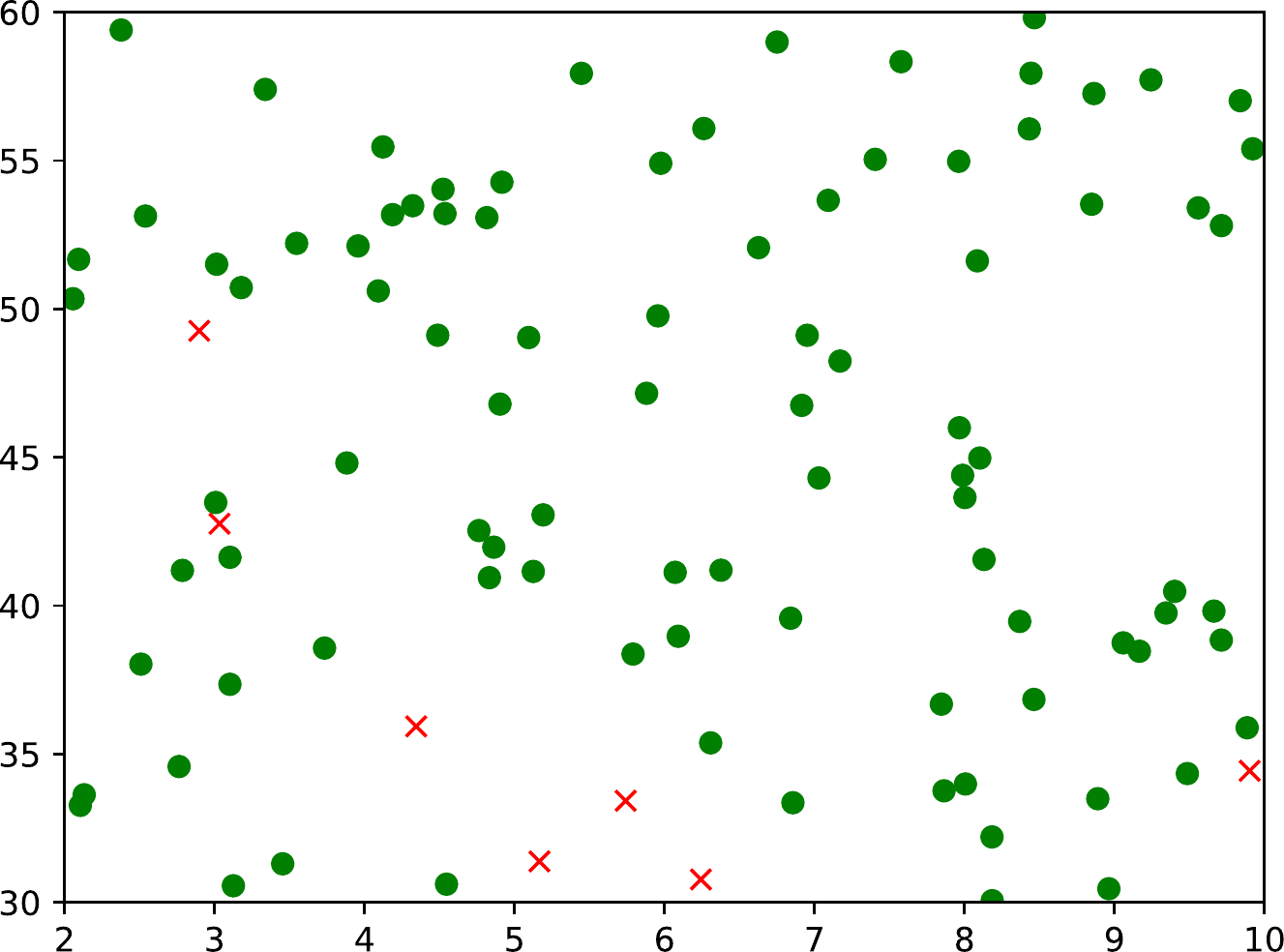}
        \caption{Random sampling (no opt.)}
        \label{fig:Pedestrian_rand_100}
    \end{subfigure}
    \hfill
    \begin{subfigure}{0.31\textwidth}
    \centering
        \includegraphics[width=0.99\textwidth]{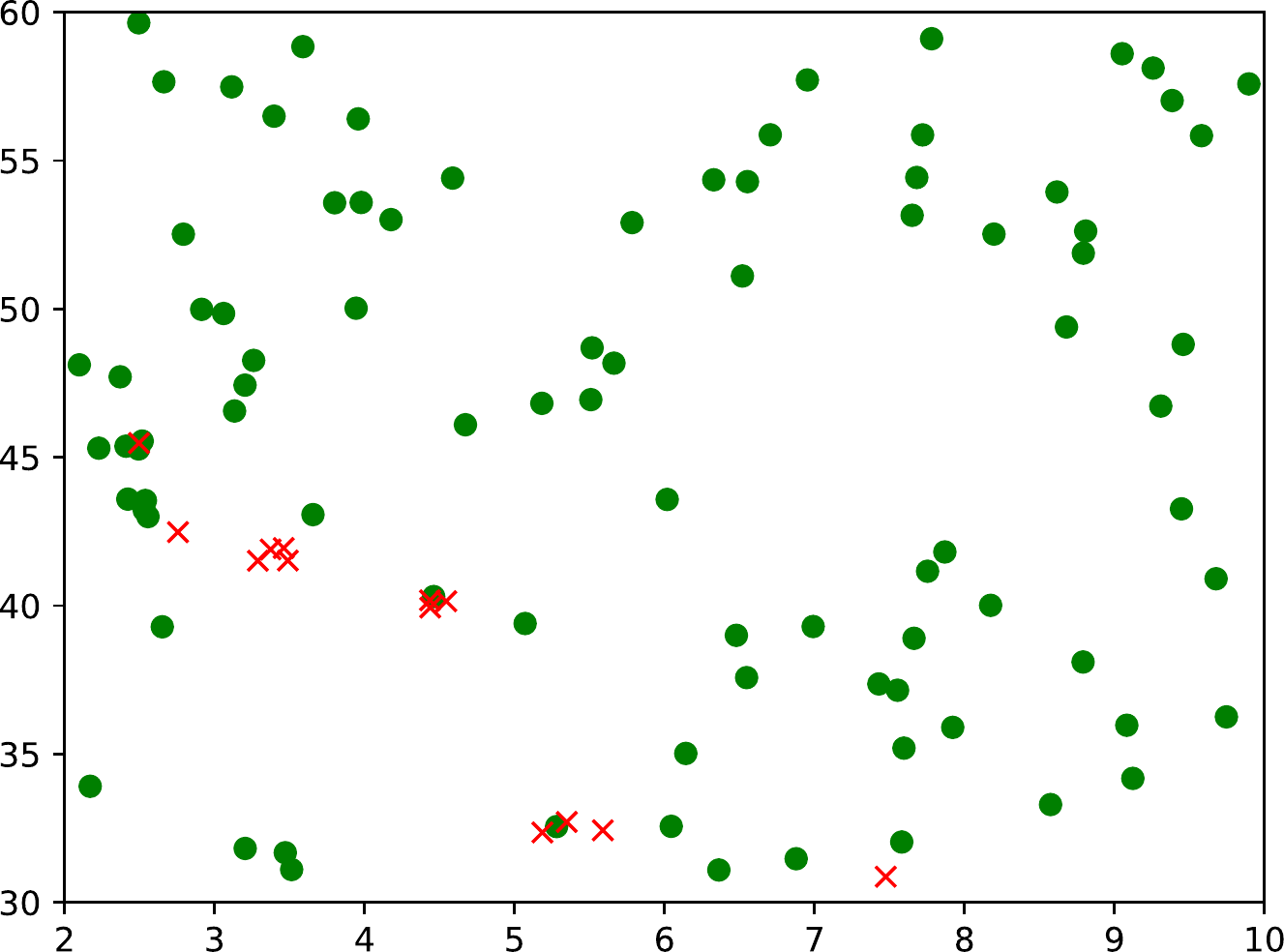}
        \caption{Random + opt. / maximizing collision.}
        \label{fig:Pedestrian_rand_sa}
    \end{subfigure}
    \hfill
    \begin{subfigure}{0.31\textwidth}
    \centering
        \includegraphics[width=0.99\textwidth]{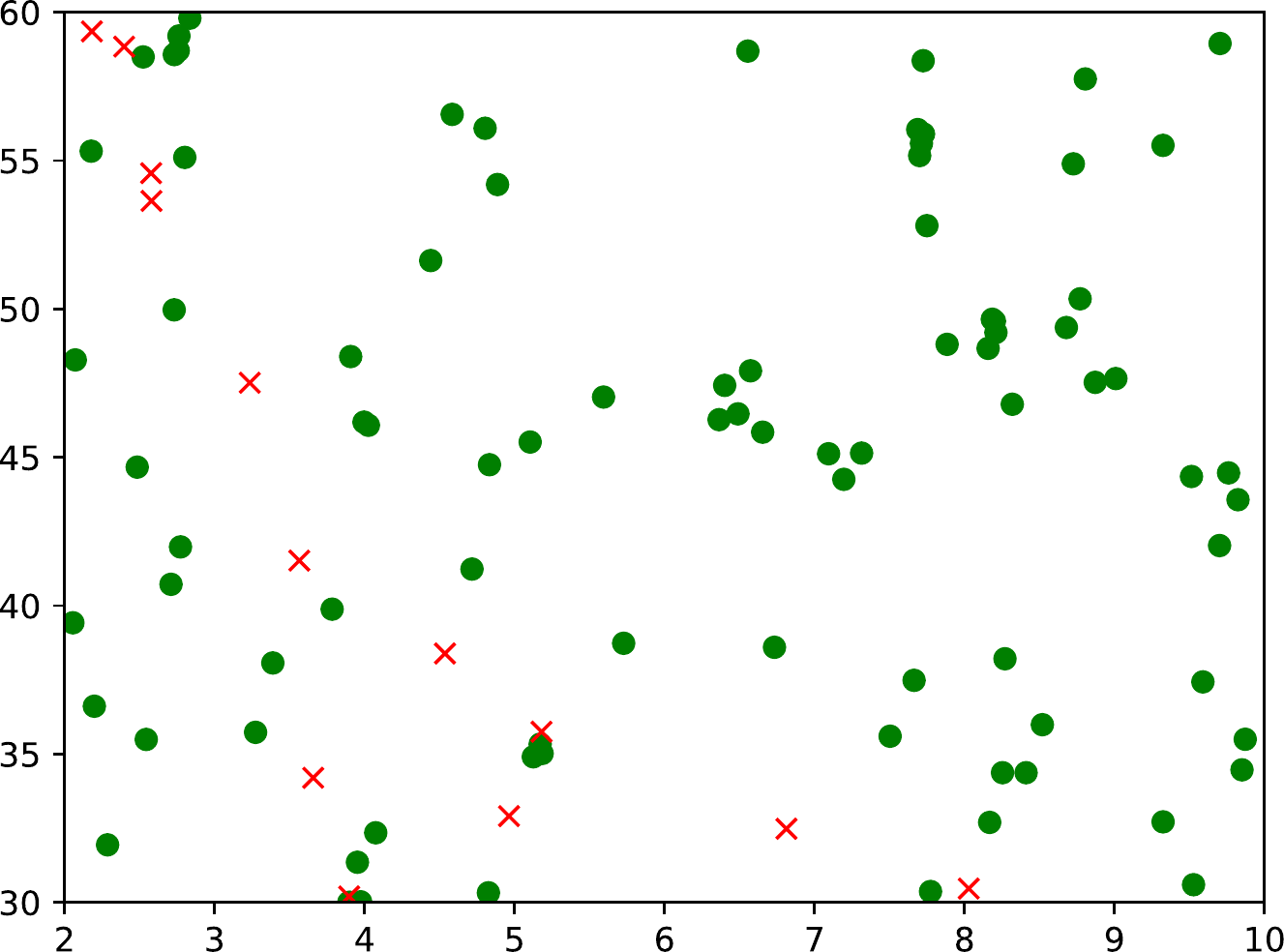}
        \caption{Random + opt. / almost failing.}
        \label{fig:Pedestrian_random_fuzzing}
    \end{subfigure}
    \begin{subfigure}{0.31\textwidth}
    \centering
        \includegraphics[width=0.99\textwidth]{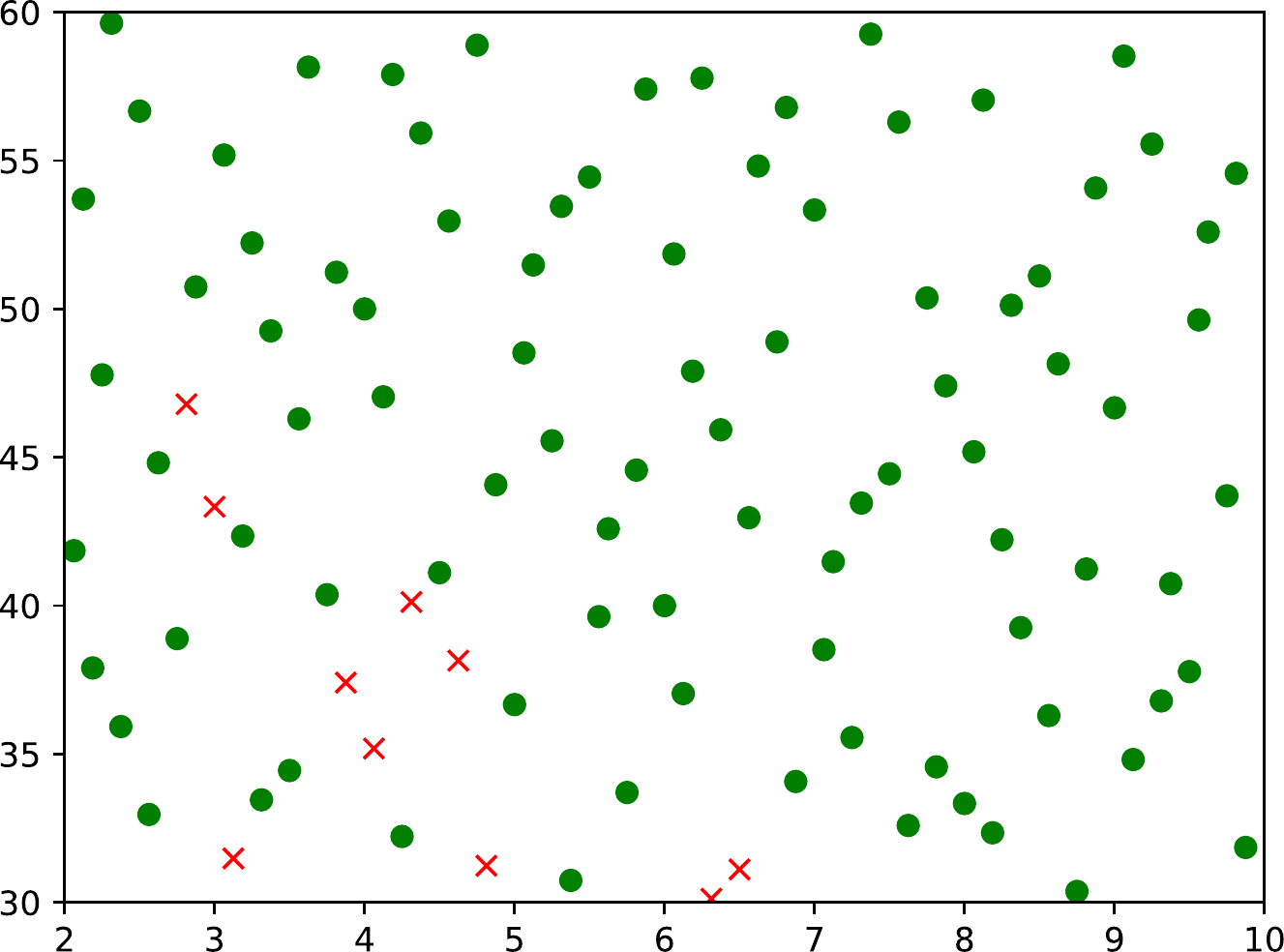}
        \caption{Halton sampling (no opt.)}
        \label{fig:Pedestrian_halton_100}
    \end{subfigure}
    \hfill
    \begin{subfigure}{0.31\textwidth}
    \centering
        \includegraphics[width=0.99\textwidth]{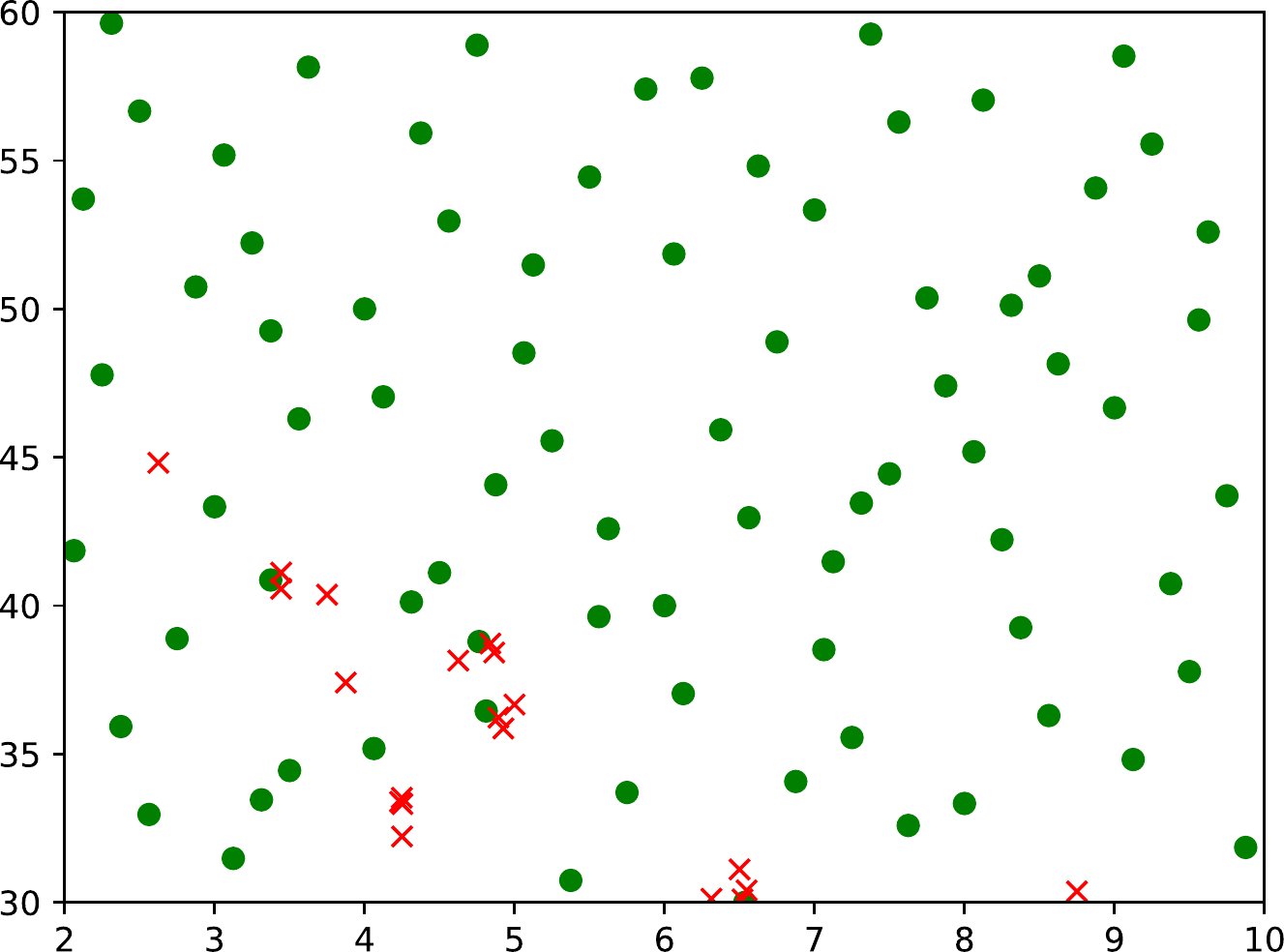}
        \caption{Halton + opt. / maximizing collision.}
        \label{fig:Pedestrian_halton_sa}
    \end{subfigure}
    \hfill
    \begin{subfigure}{0.31\textwidth}
    \centering
        \includegraphics[width=0.99\textwidth]{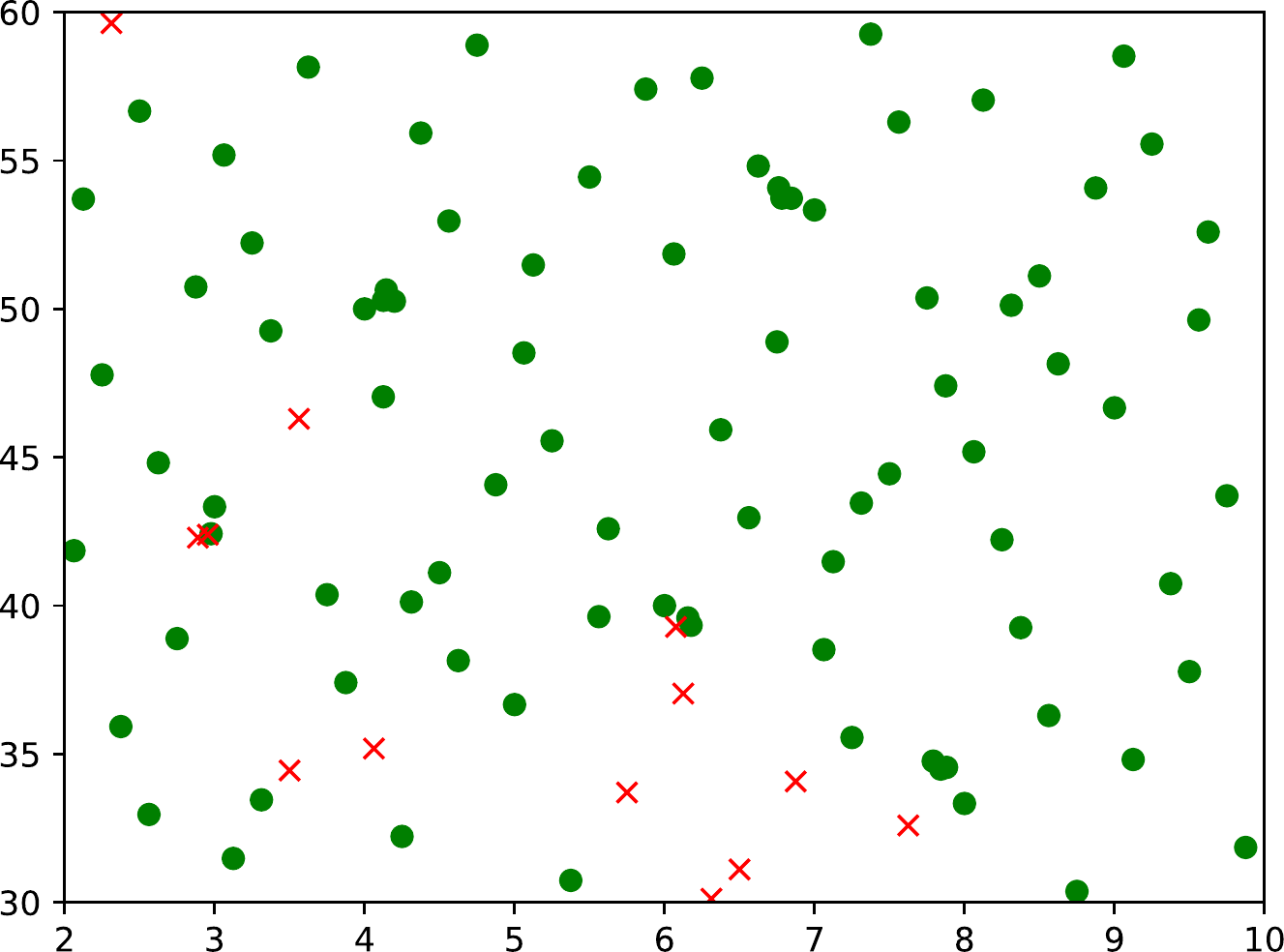}
        \caption{Halton + opt. / almost failing.}
        \label{fig:Pedestrian_halton_fuzzing}
    \end{subfigure}
    \caption{A comparison of the various test generation strategies for the jaywalking pedestrian case study. 
        The X-axis is the walking speed of the pedestrian (2 to 10 m/s).
        The Y-axis is the distance from the car when the pedestrian starts crossing (30 to 60 m).
        Passing tests are labelled with a green dot.
        Failing tests (tests with a collision) are marked with a red cross.
    }\label{fig:PedestrianJaywalking}
\end{figure}

\paragraph{Adaptive Cruise Control.}
\begin{figure}[t]
\centering
    \begin{subfigure}[b]{0.31\textwidth}
    \centering
        \includegraphics[width=0.99\textwidth]{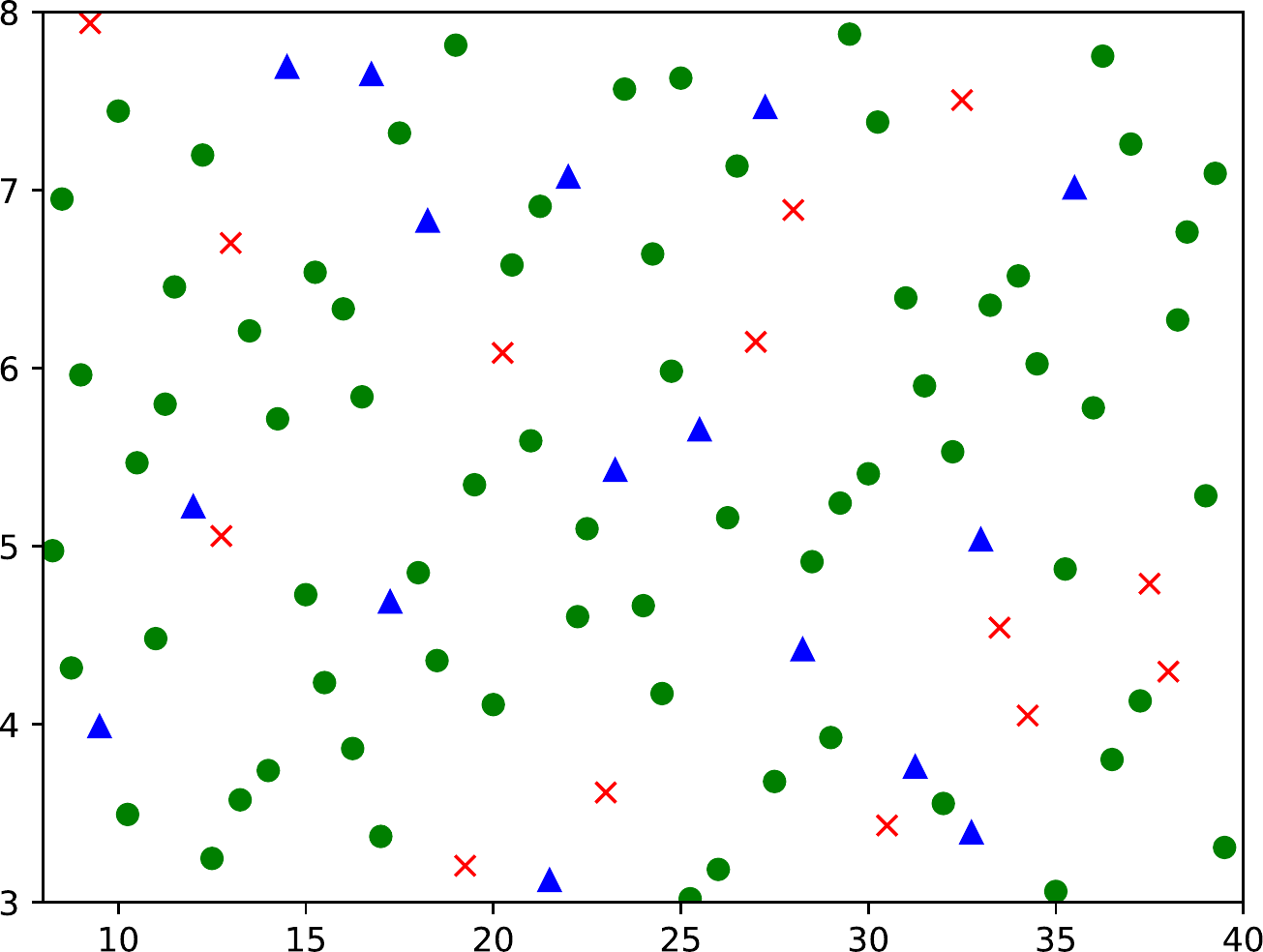}
        \caption{Initial offset (X-axis) vs. max. speed (Y-axis).}
        \label{fig:ACC_OffsetvSpeed}
    \end{subfigure}
    \hfill
    \begin{subfigure}[b]{0.31\textwidth}
    \centering
        \includegraphics[width=0.99\textwidth]{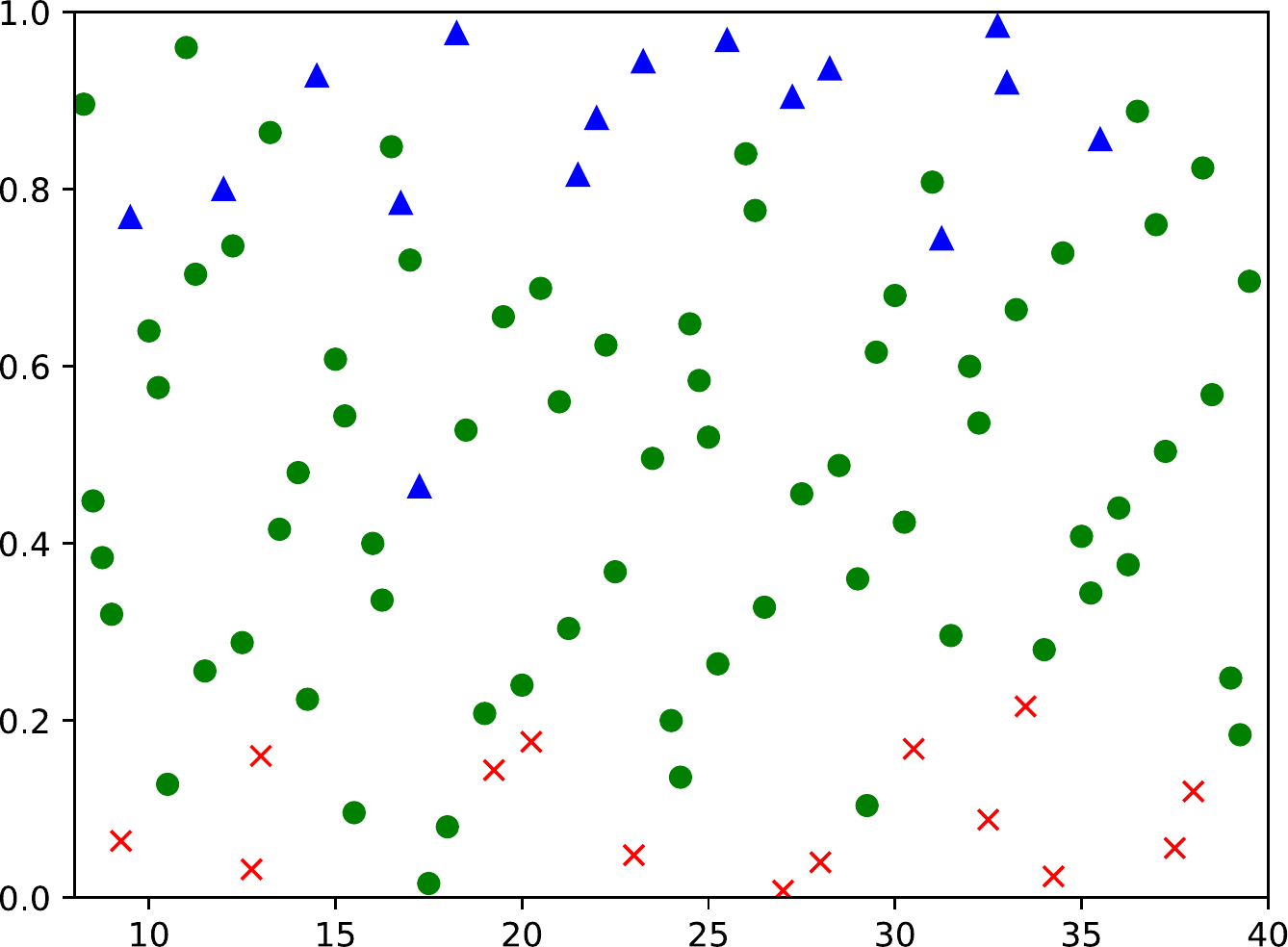}
        \caption{Initial offset (X-axis) vs. fog density (Y-axis).}
        \label{fig:ACC_OffsetvFog}
    \end{subfigure}
    \hfill
    \begin{subfigure}[b]{0.31\textwidth}
    \centering
        \includegraphics[width=0.99\textwidth]{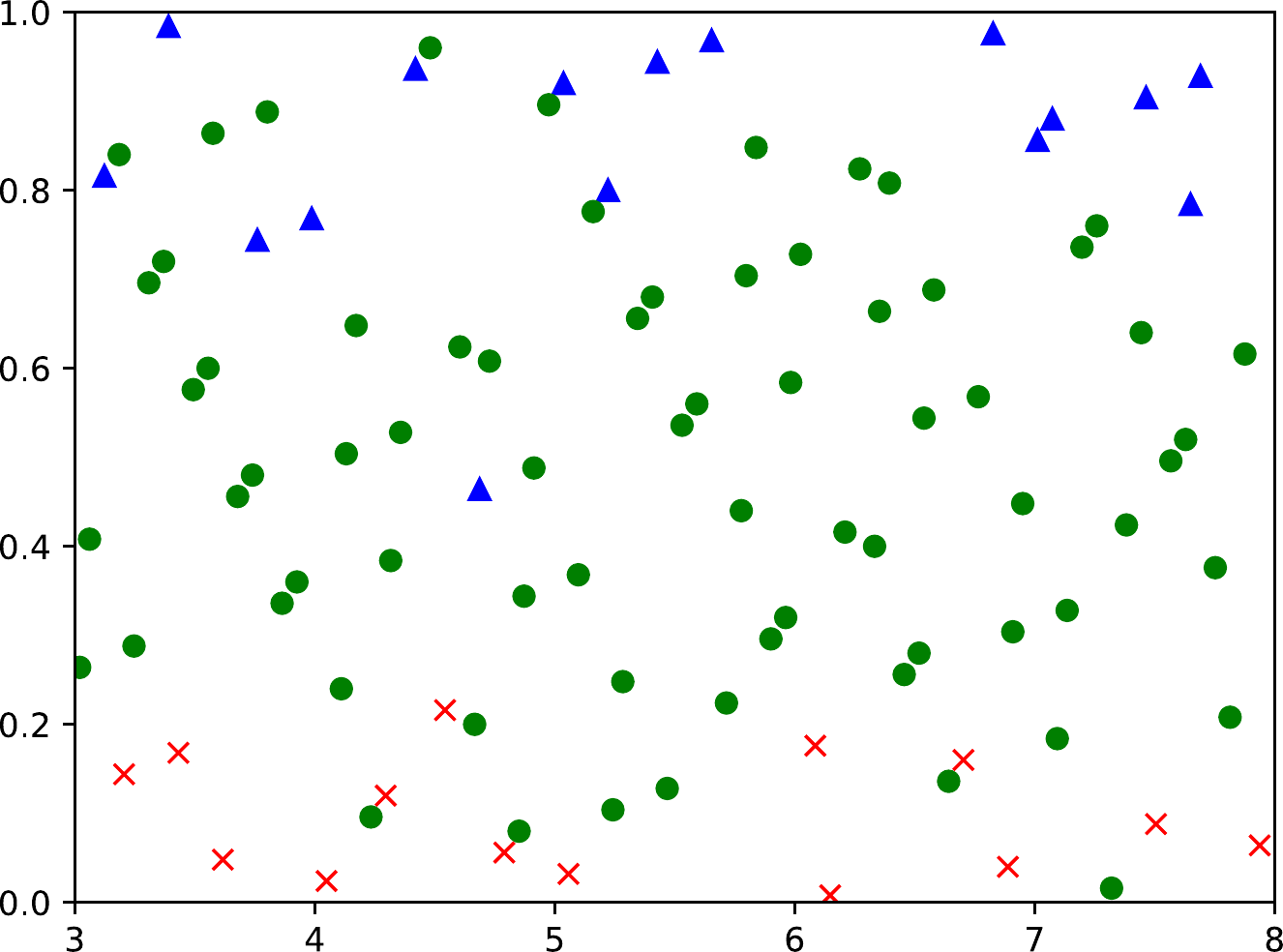}
        \caption{Max. speed (X-axis) vs. fog density (Y-axis).}
        \label{fig:ACC_SpeedvFog}
    \end{subfigure}
    \caption{Continuous test parameters of the Adaptive Cruise Control study plotted against each other: the initial offset of the lead car (8 to 40 m), the lead car's maximum speed (3 to 8 m/s) and the fog density (0 to 1). 
    Green dots, red crosses, and blue triangles denote passing tests, collisions, and inactivity respectively.
    }
    \label{fig:ACC}
\end{figure}
\begin{table}[t]
\centering
\small
\caption{Parameterized test on Adaptive Cruise Control,
separated for each value of discrete parameters, and low and high values of continuous parameters.
A test \emph{passes} if there are no collisions and no inactivity (the overall distance moved by the test vehicle is more than 5 m.
The average offset (in m) maintained by the test vehicle to the lead car (for passing tests) is also presented. }
\label{tbl:acc_results}
\begin{tabular}{ l c c c c c c c c c c c c }
\toprule
 & \multicolumn{6}{c}{Discrete parameters} & \multicolumn{6}{c}{Continuous parameters}\\
\midrule
         & \multicolumn{2}{c}{Num. lanes} & \multicolumn{4}{c}{Lead car color} & \multicolumn{2}{c}{Initial offset (m)} & \multicolumn{2}{c}{Speed (m/s)} & \multicolumn{2}{c}{Fog density}\\
\cmidrule(r){2-3}\cmidrule(r){4-7} \cmidrule(r){8-9}\cmidrule(r){10-11}\cmidrule(r){12-13}
 & 2 & 4 & Black & Red & Yellow & Blue   & $< 24$ & $\geq 24$ & $< 5.5$ & $\geq 5.5$ & $< 0.5$ & $\geq 0.5$  \\ \midrule
 Test iters & 54 & 46 & 24 & 22 & 27 & 27 & 51 & 49 & 52 & 48 & 51 & 49\\
 Collisions & 7 & 7 & 3 & 3 & 6 & 2 & 6 & 8 & 8 & 6 & 12 & 0\\
 Inactivity & 12 & 4 & 4 & 4 & 6 & 2 & 9 & 7 & 9 & 7 & 1 & 15\\
 Offset (m) & 42.4 & 43.4 & 46.5 & 48.1 & 39.6 & 39.1 & 33.7 & 52.7 & 38.4 & 47.4 & 36.5 & 49.8\\
 \bottomrule
\end{tabular}
\end{table}
We now create and test in an environment with our test vehicle following a car (lead car) on the same lane. 
The lead car's behavior is programmed to drive on the same lane as the test vehicle, with a certain maximum speed.
This is a very typical driving scenario that engineers test their implementations on.
We use $5$ test parameters: the initial lead of the lead car to the test vehicle ($[8 \: m,\:  40 \: m]$), the lead car's maximum speed ($[3 \: m/s,\: 8 \: m/s]$),
density of fog\footnote{0 denotes no fog and 1 denotes very dense fog (exponential squared scale).} in the environment ($[0,1]$), number of lanes on the road ($\{2,\:  4\}$), and color of the lead car ($\{Black, \: Red, \: Yello, \: Blue\}$).
We use both, \lstinline{CollisionMonitor}\footnote{the monitor additionally calculates the mean distance of the test vehicle to the lead car during the test, which is used for later analysis.} and \lstinline{DistanceMonitor}, as presented in Section~\ref{sec:LangThroughExamples}.
A test \emph{passes} if there is no collision and the autonomous vehicle moves atleast 5 m during the simulation duration (15 s).

We use \tool's default test generation strategy, i.e., Halton sampling for continuous parameters and Random sampling for discrete parameters (no optimization or fuzzing).
The SUT is the same CNN as in the previous case study. 
It is trained on 1034 training samples, which are obtained by manually driving behind a red lead car on the same lane of a 2-lane road with the same maximum velocity (5.5 m/s) and no fog.

The results of this case study are presented in Table~\ref{tbl:acc_results}.
Looking at the discrete parameters, the number of lanes does not seem to contribute towards a risk of collision.
Surprisingly, though the training only involves a Red lead car, the results appear to be the best for a Blue lead car.
Moving on to the continuous parameters, the fog density appears to have the most significant impact on test failures (collision or vehicle inactivity).
In the presence of dense fog, the SUT behaves pessimistically and does not accelerate much (thereby causing a failure due to inactivity).
These are all interesting and useful metrics about the performance of our SUT.
Plots of the results projected on to continuous parameters are presented in Figure~\ref{fig:ACC}.

\subsection{Results and Analysis}
\label{sec:CaseStudies_Analysis}
We now summarize the results of our evaluation with respect to our \textbf{RQ}s:\\
\textbf{RQ 1}:
All the three case studies involve varied, rich and dynamic environments.
They are representative of tests engineers would typically want to do, and we parameterize many different aspects of the world and the dynamic behavior of its components.
These designs are at most $70$ lines of code.
This provides confidence in \tool's ability of providing an easy interface for the design of realistic test environments.
\\
\textbf{RQ 2}:
Our default test generation strategies are found to be quite effective at exploring the parameter space systematically, eliminating large unexplored gaps, and at the same time, successfully identifying problematic cases in all the three case studies.
The jaywalking pedestrian study demonstrates that optimization and local search are possible on top of these strategies, and are quite effective in finding the relevant scenarios.
The adaptive cruise control study tests over $5$ parameters, which is more than most related works, and even guarantees good coverage of this parameter space.
Therefore, it is amply clear that \tool's test input generation methods are useful and effective.
\\
\textbf{RQ 3}:
The road segmentation case study uses a well-performing neural network for object segmentation, and we are able to detect degraded performance for automatically generated test inputs.
Whereas this study focuses on static image classification, the next two, i.e., the jaywalking pedestrian and the adaptive cruise control study uncover poor performance on simulated driving, using a popular neural network architecture for self driving cars.
Therefore, we can safely conclude that \tool can find bugs in various different kinds of systems related to autonomous driving.

\subsection{Threats to Validity}
\label{sec:threats-validity}

The \emph{internal validity} of our experiments depends on having implemented our system correctly and, more importantly, trained and used the neural networks
considered in the case studies correctly.
For training the networks, we followed the available documentation and inspected our examples to ensure that we use an appropriate training procedure.
We watched some test runs and replays of tests we did not understand. 
Furthermore, our implementation logs events and we also capture images, which allow us to check a large number of tests.

In terms of threats to external validity, the biggest challenge in this project has been finding systems that we can easily train and test in complex driving scenarios.
Publicly available systems have limited capabilities and tend to be brittle.
Many networks trained on real world data do not work well in simulation.
We therefore re-train these networks in simulation.
An alternative is to run fewer tests, but use more expensive and visually realistic simulations.
Our test generation strategy maximizes coverage, even when only a few test iterations can be performed due to high simulation cost.

\section{Related Work}
\label{sec:RelatedWork}

Traditionally, test-driven software development paradigms \cite{tdd} have advocated testing and mocking frameworks to test software early and often.
Mocking frameworks and mock objects \cite{Mackinnon00endo-testing:unit,Mockito} allow programmers to test a piece of code against an API specification. 
Typically, mock objects are stubs providing outputs to explicitly provided lists of inputs of simple types, 
with little functionality of the actual code.
Thus, they fall short of providing a rich environment for autonomous driving.
\tool can be seen as a mocking framework for reactive, physical systems embedded in the 3D world.
Our notion of constraining streams is inspired by work on declarative mocking \cite{Millstein}.

\paragraph{Testing Cyber-Physical Systems.}
There is a large body of work on automated test generation tools for cyber-physical systems
through heuristic search of a high-dimensional continuous state space.
While much of this work has focused on low-level controller interfaces
\cite{STaliro,STaliro-auto,Donze2010,deshmukh2015stochastic,Sankaranarayanan2012,DeshmukhBayesian17}
rather than the system level, specification and test generation techniques arising from this work---for example,
the use of metric and signal temporal logics or search heuristics---can be adapted to our setting.
More recently, test generation tools have started targeting autonomous systems
under a simulation-based semantic testing framework similar to ours.
In most of these works, visual scenarios are either fixed by hand \cite{DreossiDS17, raja_ase, raja_icse, seshiapldi19, toyotaSim-basedAdversarial, tuncali2019requirements,test_harness_sim,DBLP:conf/issta/GambiMF19},
or are constrained due to the model or coverage criteria \cite{situation_coverage, sys_testing, o2018scalable}.
These analyses are shown to be preferable to the application of random noise on the input vector.
Additionally, a simulation-based approach filters benign misclassifications from misclassifications that actually lead to bad or dangerous behavior.
Our work extends this line of work and provides an expressive language to design parameterized environments and tests.
\textsc{AsFault} \cite{DBLP:conf/issta/GambiMF19} uses random search and mutation for procedural generation of road networks for testing.
\textsc{AC3R} \cite{DBLP:conf/sigsoft/GambiHF19} reconstructs test cases from accident reports.



To address problems of high time and infrastructure cost of testing autonomous systems, several simulators have been developed.
The most popular is Gazebo \cite{url_GazeboCarSim} for the ROS \cite{ROS} robotics framework.
It offers a modular and extensible architecture, however falls behind on visual realism and complexity of environments that can be generated with it.
To counter this, game engines are used. 
Popular examples are TORCS \cite{url_torcs}, CARLA \cite{carla17}, and AirSim \cite{airsim17}
Modern game engines and support creation of realistic urban environments.
Though they enable visually realistic simulations and enable detection of infractions such as collisions, the environments themselves are difficult to design.
Designing a custom environment involves manual placement of road segments, buildings, and actors (as well as their properties).
Performing many systematic tests is therefore time-consuming and difficult.
While these systems and \tool share the same aims and much of the same infrastructure, \tool focuses on procedural design and systematic testing, backed by a relevant coverage criteria.

\paragraph{Adversarial Testing.}
Adversarial examples for neural networks \cite{Goodfellow:2014,Szegedy:2013}
introduce perturbations to inputs that cause a
classifier to classify ``perceptually identical'' inputs differently.
Much work has focused on finding adversarial examples in the context of autonomous driving
as well as on training a network to be robust to perturbations 
\cite{Papernot:2017,Song:2017,DBLP:conf/tacas/WickerHK18,Vechev:2018,AI2Robust18}.
Tools such as \textsc{DeepXplore} \cite{DBLP:conf/sosp/PeiCYJ17}, \textsc{DeepTest} \cite{deeptestICSE}, \textsc{DeepGauge} \cite{Ma:2018:DMT:3238147.3238202}, and SADL \cite{Kim:2019:GDL:3339505.3339634} define a notion of coverage for neural networks based on the number of neurons activated during tests compared against the total number of neurons in the network and activation during training.
However, these techniques focus mostly on individual classification tasks and apply 2D transformations on images.
In comparison, we consider the closed-loop behavior of the system and our parameters directly change the world rather than apply transformations post facto.
We can observe, over time, that certain vehicles are not detected, which is more useful to testers than a single misclassification \cite{jens_sbt_eval}.
Furthermore, it is already known that structural coverage criteria may not be an effective strategy for finding errors in classification \cite{DBLP:conf/icse/LiM0C19}.
We use coverage metrics on the test space, rather than the structure of the neural network.
Alternately, there are recent techniques to verify controllers implemented as neural networks through constraint solving or abstract interpretation
\cite{AI2Robust18,DBLP:conf/cav/HuangKWW17,DBLP:conf/ijcai/RuanHK18,DBLP:conf/tacas/WickerHK18,DBLP:conf/hybrid/DuttaCJST19}.
While these tools do not focus on the problem of autonomous driving, their underlying techniques can be combined
in the test generation phase for \tool.

%


\section{Future Work and Conclusion}

Deploying autonomous systems like self-driving cars in urban environments raises several safety challenges.
The complex software stack processes sensor data, builds a semantic model of the surrounding world, makes decisions, plans trajectories, and controls the car.
The end-to-end testing of such systems requires the creation and simulation of whole worlds, with different tests representing different world and parameter configurations. 
\tool tackles these problems by
\begin{inparaenum}[(i)]
\item enabling procedural construction of diverse scenarios, with precise control over elements like layout roads, physical and visual properties of objects, and behaviors of actors in the system, and
\item using quasi-random testing to obtain good coverage over large parameter spaces.
\end{inparaenum}

In our evaluation, we show that \tool enables easy design of environmnents and automated testing of autonomous agents implemented using neural networks.
While finding errors in sensing can be done with only a few static images, we show that \tool also enables the creation of longer 
test scenarios which exercise the controller's feedback on the environment.
Our case studies focused on \emph{qualitative} state space exploration.
In future work, we shall perform \emph{quantitative} statistical analysis to understand the sensitivity
of autonomous vehicle behavior on individual parameters.

In the future, we plan to extend \tool's testing infrastructure to also aid in the training of deep neural networks that require large amounts of high quality training data.
For instance, we show that small variations in the environment result in widely different results for road segmentation.
Generating data is a time consuming and expensive task.
\tool can easily generate labelled data for static images.
For driving scenarios, we can record a user manually driving in a parameterized \tool environment and augment this data by varying parameters that should not impact the car's behavior.
For instance, we can vary the color of other cars, positions of pedestrians who are not crossing, or even the light conditions and sensor properties (within reasonable limits).

\subsubsection*{Acknowledgements}
This research was funded in part by the Deutsche Forschungsgemeinschaft project \textsc{389792660-TRR 248}
and by the European Research Council under the
Grant Agreement \textsc{610150} (ERC Synergy Grant ImPACT).

\bibliographystyle{splncs04}
\bibliography{Bibliography}

\clearpage
\appendix
In these Appendices to the main paper, we present additional case studies performed using \tool.
We also provide additional description and a code sample for connection and composition of road elements, and a sample output to \opendrive \cite{opendrive}.

\section{Testing networks trained on standard datasets}
\label{sec:pre_trained}
Many autonomous driving systems have on-board components for computer vision tasks like road segmentation, traffic light and traffic sign classification, vehicle detection, 
and optical flow.
In the following tests, instead of training the SUT (a deep neural network) inside our simulation environment, we test components trained on real 
world datasets.
We present results for road detection and vehicle detection.

\paragraph{Test environment.}
For the tests, we designed a highly parameterized environment using \tool's programmatic interface.
The environment consisted of 4 \lstinline{StraightRoadSegment}s connected by a \lstinline{CrossIntersection}.

\begin{figure}
    \centering
    \begin{subfigure}{0.55\textwidth}
        \centering
        \includegraphics[width=\textwidth]{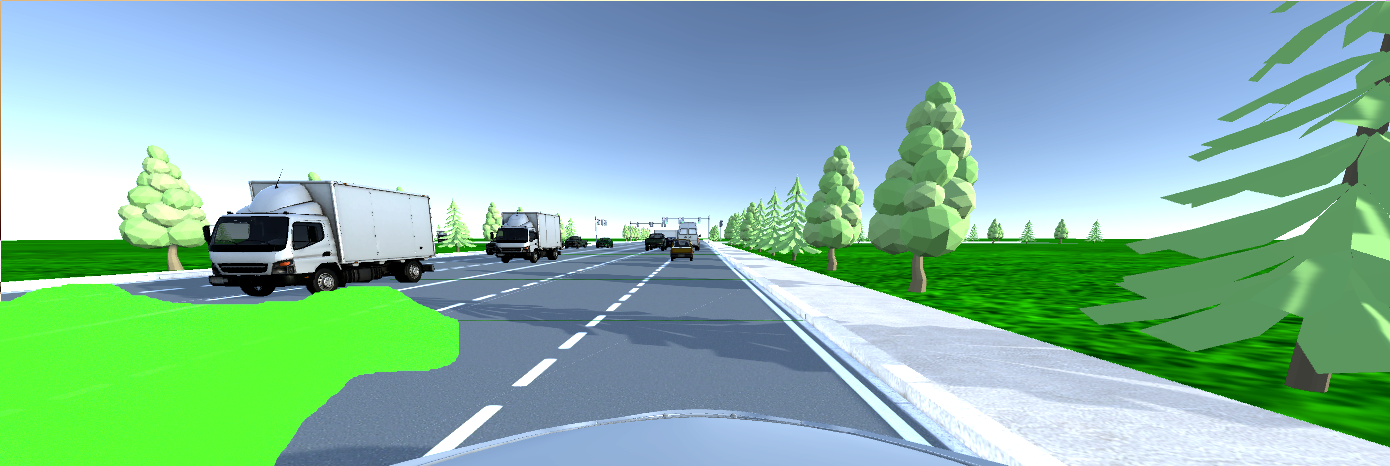}
        \label{fig:focal10}
    \end{subfigure}  
    
    \begin{subfigure}{0.55\textwidth}
        \centering
        \includegraphics[width=\textwidth]{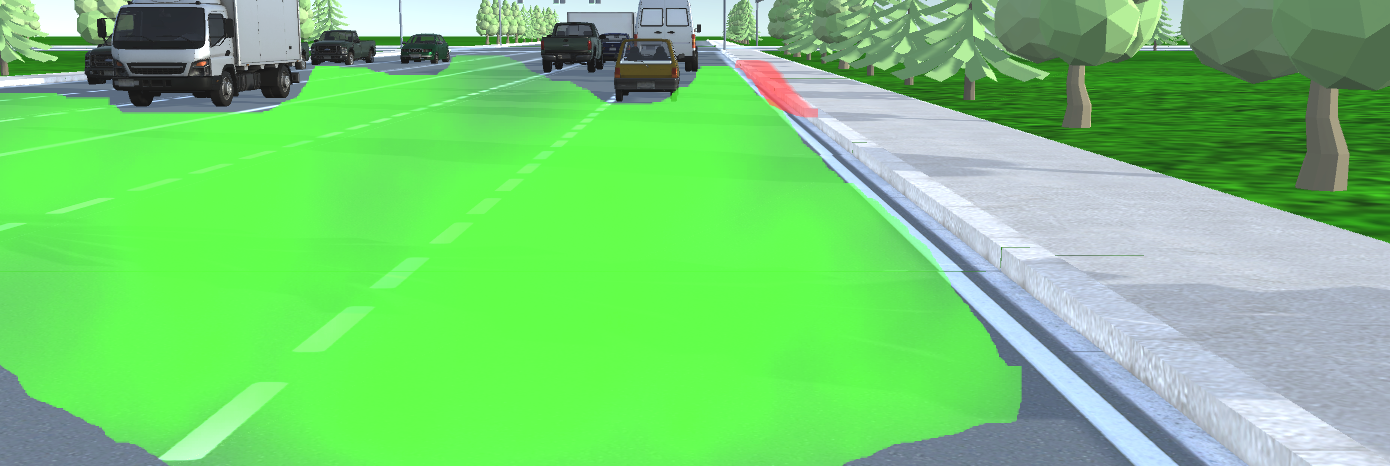}
        \label{fig:focal34}
    \end{subfigure}
    \caption{Sample output from a road detection system. 
    Green pixels represent correctly identified road, red pixels represent pixels incorrectly identified as road.    
    A longer focal length (34mm) results in better road detection in comparison to a shorter focal length (10mm).}
    \label{fig:focal_width_rd}
\end{figure}

The test has three discrete parameters and three continuous parameters:
\begin{inparaenum}[(i)]
\item The number of lanes is either $2$ or $4$ (discrete).
\item The light condition corresponds to a morning, noon, or an evening drive (discrete).
\item The number of other cars on the road ranges from $2$ to $9$ (discrete).
\item The camera focal length is in the $[18,22]$ mm interval (continuous).
\item The height of the mounting point of the camera varies from $1.9$m to $2.2$m (continuous).
\item Finally, the camera looks slightly down with a pitch angle between $-10$ and $-12$ degrees (continuous).
\end{inparaenum}
Many of our parameters correspond to the vehicle's camera. 
These were chosen because in preliminary tests, small perturbations to the camera's properties led to drastically different 
results (see Figure~\ref{fig:focal_width_rd}).
We perform 100 test iterations using \tool's default test generation scheme.

\begin{figure}[t]
    \centering
    \includegraphics[width=0.55\textwidth]{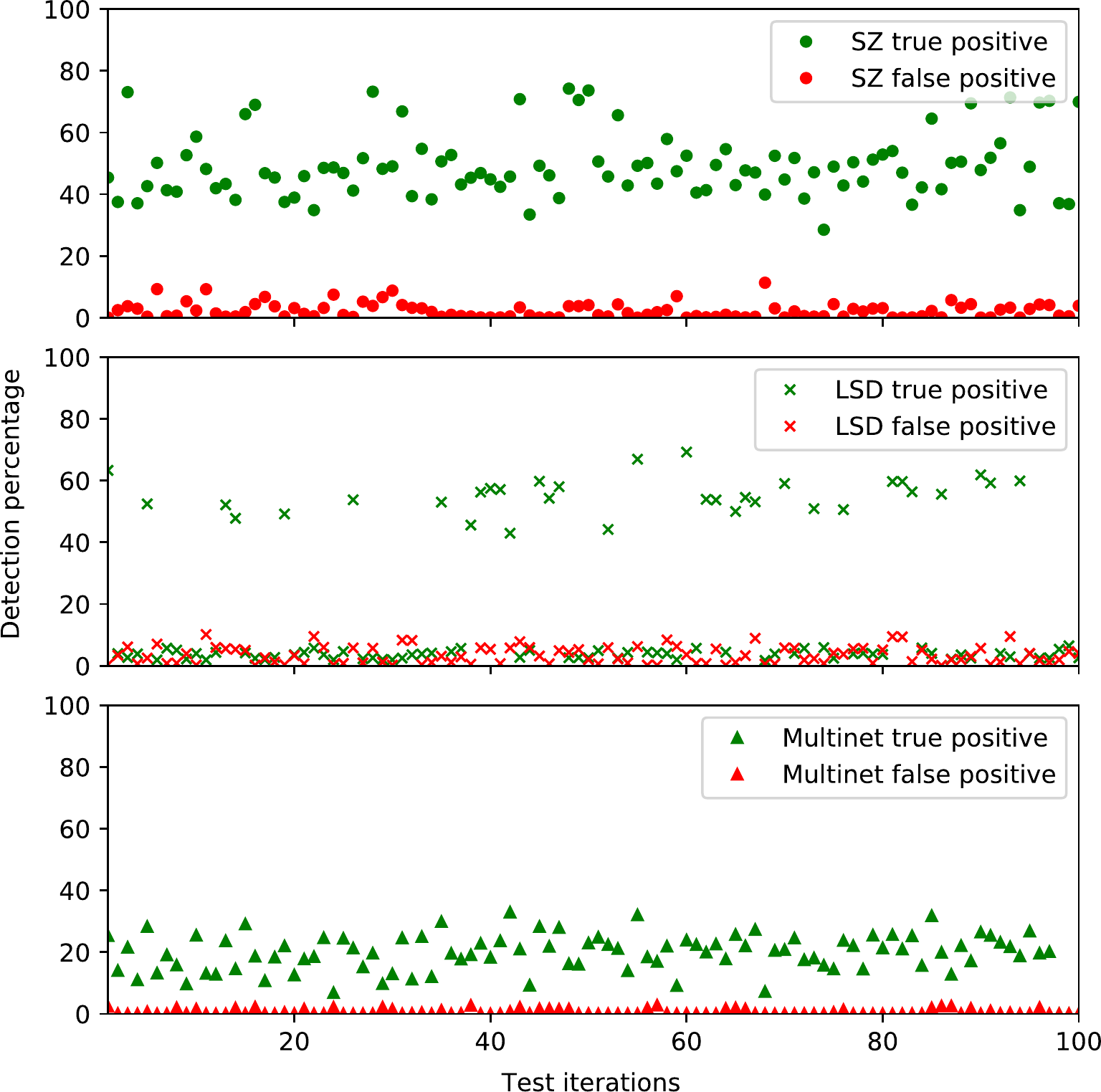}
    \caption{Road segmentation results (\% of the ground truth).}
    \label{fig:focal_graph}
\end{figure}

\paragraph{Road segmentation.}
The SUTs here take RGB images as input and return those pixels that are estimated to be a part of the road.
We tested:
\begin{inparaenum}[(i)]
\item the convolutional neural network from Simonyan and Zisserman (popular as VGGNet) \cite{Simonyan14verydeep},
\item Multinet from Teichmann, Weber et al. \cite{TeichmannWZCU16}, a top performer on the KITTI Road Estimation Benchmark, and
\item the fully convolutional network by Long, Shelhamer and Darrell \cite{LongSD14},
\end{inparaenum}
All three are trained on the KITTI road segmentation dataset \cite{kitti2013} (289 images).
The first and third networks do not have a name, so we use initials of the authors' names, SZ and LSD, respectively.
Figure~\ref{fig:focal_graph} shows the results for the 100 test iterations ($x$-axis).
We plot results in the order in which the tests were performed.
The $y$-axis shows the percentage of the ``ground truth'' road identified as road by the method.
A cursory look did not reveal any correlation between road segmentation performance and parameter choice.
We observe that SZ is the best performer overall.
What is quite striking is the results of LSD:
in these tests, it either performs well, or not at all.
Except for the poorly performing examples of LSD, false positives are not an issue, generally.
Our hypothesis is that the networks do not generalize sufficiently from the limited training data and 
images that are too different from training lead to poor results.

\begin{figure}[t]
    \centering
    \includegraphics[width=0.55\textwidth]{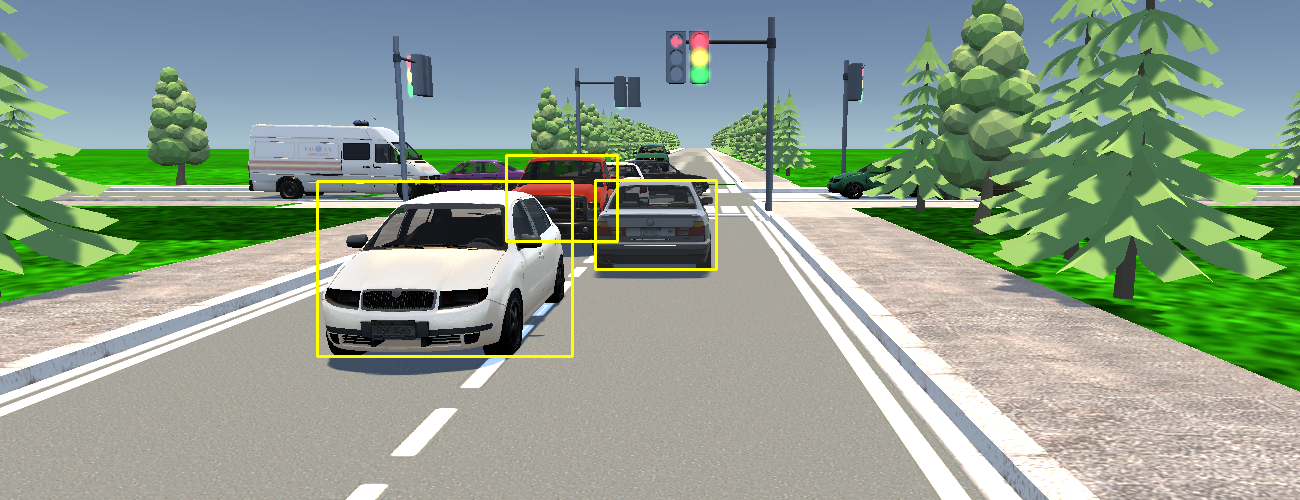}\\[1ex]

    \includegraphics[width=0.55\textwidth]{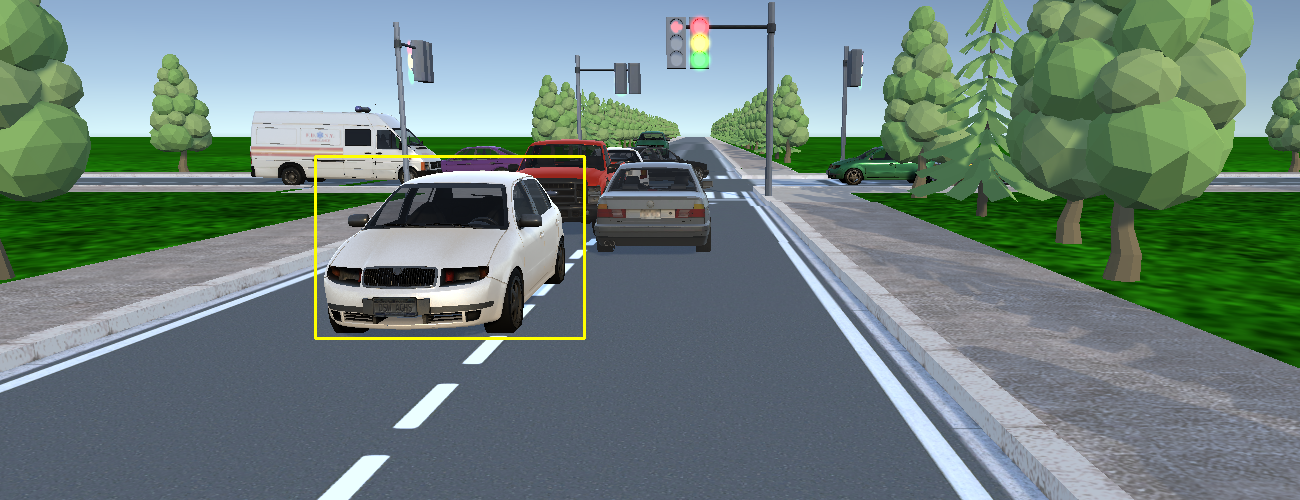}
    \caption{Sample outputs of a vehicle detection system.
             Small changes to the environmental condition lead to missed cars.}
    \label{fig:car_detection}
\end{figure}

\begin{figure}
    \centering
    \includegraphics[width=0.55\textwidth]{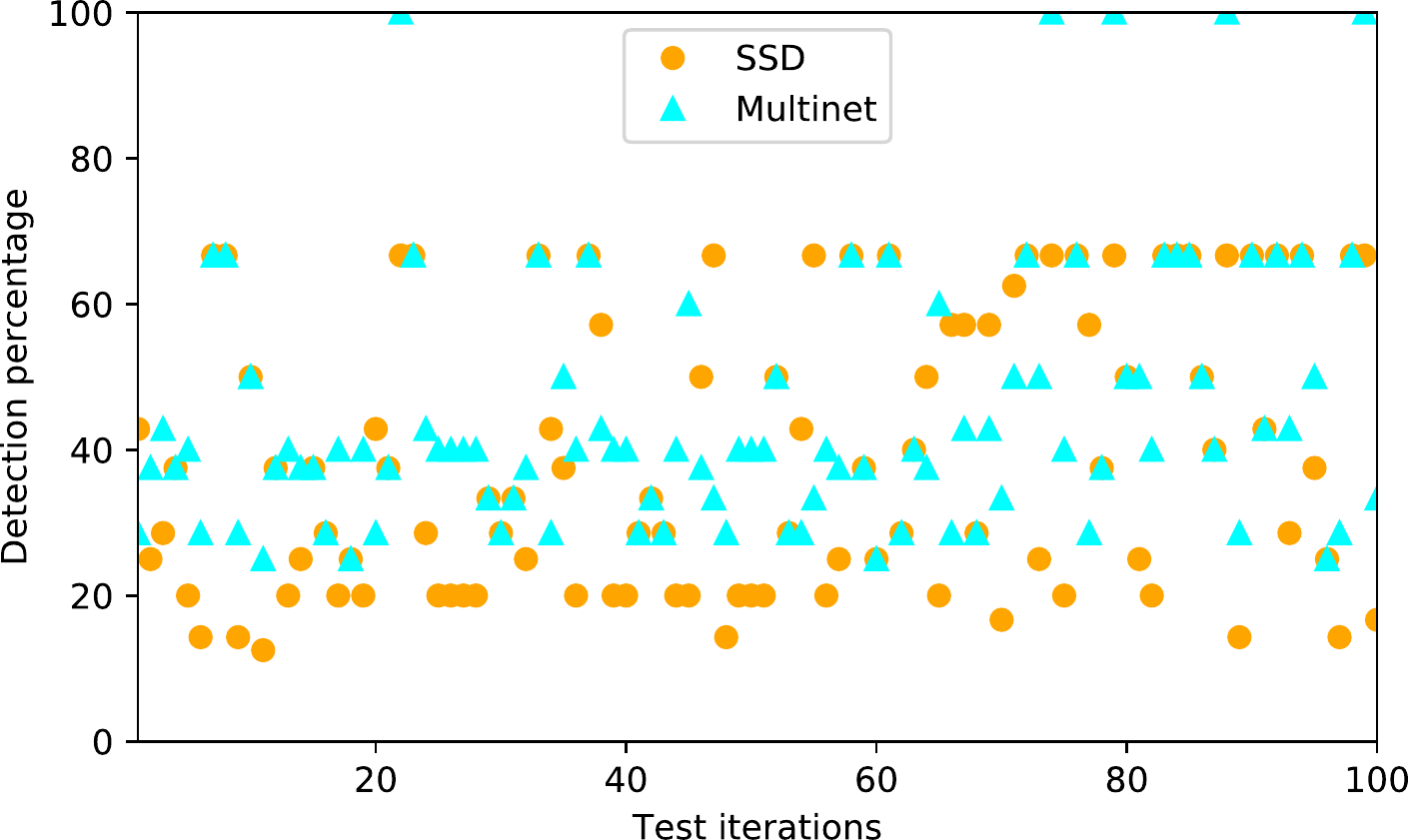}
    \caption{Vehicle detection rates for the two SUTs.}
    \label{fig:vd}
\end{figure}

\paragraph{Vehicle detection.}
The SUTs here take RGB images as input and return bounding boxes around pixels that correspond to vehicles.
Figure \ref{fig:car_detection} shows an example of a vehicle detection system's output.
To detect other vehicles in the vicinity of the autonomous car, we used:
\begin{inparaenum}[(i)]
\item the single shot multibox detector (SSD), a deep neural network \cite{vehicleDetectionSSD2015}, trained with the Pascal Object Recognition Database Collection \cite{pascal-voc-2011},
\item Multinet \cite{TeichmannWZCU16} like in the previous experiment.
\end{inparaenum}
Figure \ref{fig:vd} summarizes the results.
The results are again in the order of the tests.
In this experiment, we did not observe any false positives.
Overall, Multinet performs better than SSD but these systems are much closer than in the previous experiment.
While the detection rates may look disappointing, factors like occlusion as seen in Figure \ref{fig:car_detection} make it difficult to detect all the cars.
The two experiments presented here highlight the fact that even with quite narrow parameter ranges (especially for the camera), the quality of results can vary widely.

\section{Additional tests on driving behavior}
\label{case_study_behaviour}

In the case studies that follow, we test the NVIDIA behavioral cloning framework \cite{nvidiaEndToEnd} to perform tests on autonomous vehicle behavior.
The SUT takes RGB images as input and returns the corresponding throttle or steering control to be applied.
The network is trained inside \tool's simulation environment and the specific training procedure is also described for each case study presented.
The primary aim of these case studies is to highlight specific testing features of \tool.

\begin{figure}
    \centering
    \begin{subfigure}[b]{0.45\textwidth}
        \includegraphics[width=\textwidth]{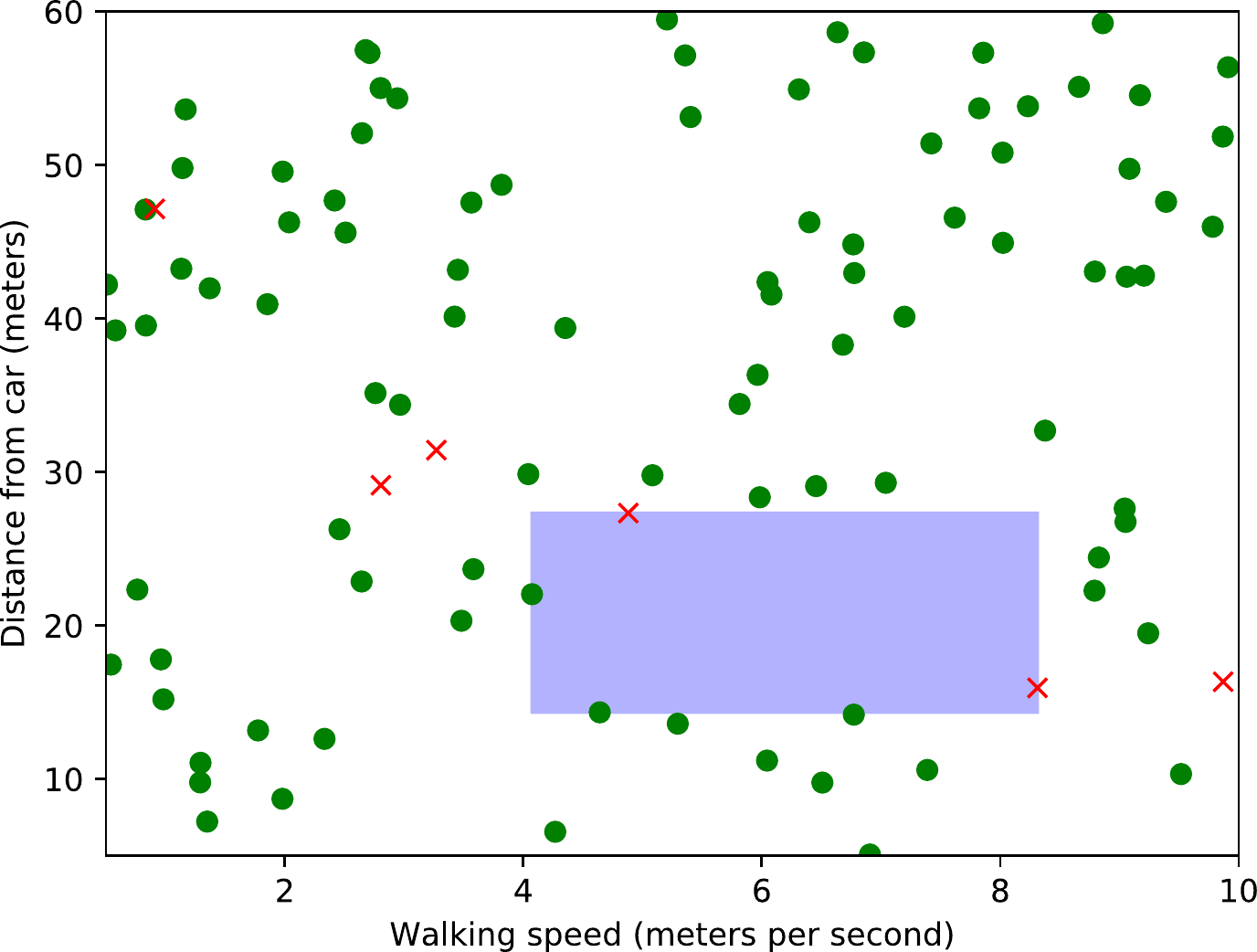}
        \caption{Random (dispersion is $0.105$)}
        \label{fig:Pedestrian_rand_100}
    \end{subfigure}
    ~ 
    \begin{subfigure}[b]{0.45\textwidth}
        \includegraphics[width=\textwidth]{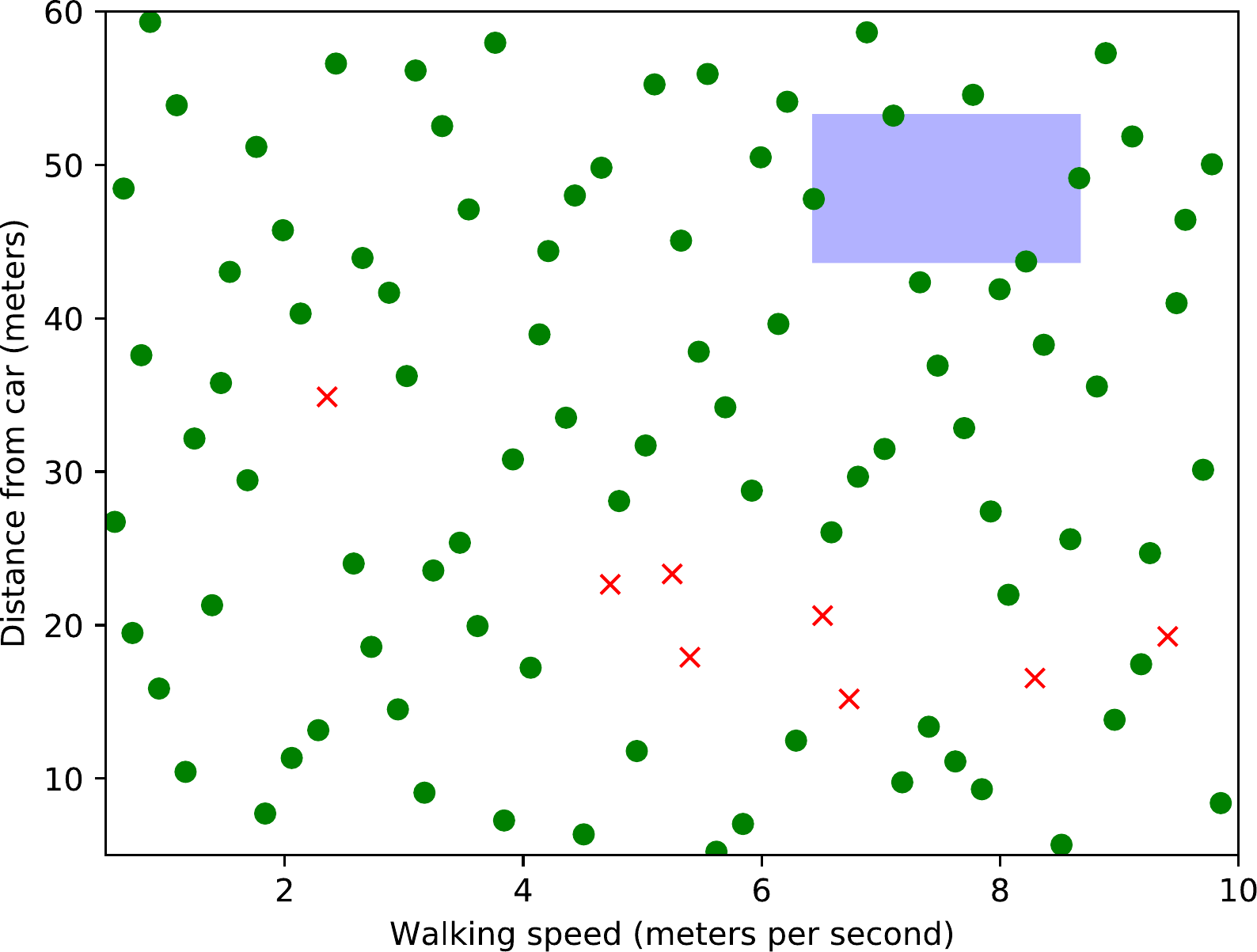}
        \caption{Halton (dispersion is $0.041$)}
        \label{fig:Pedestrian_halton_100}
    \end{subfigure}
    \caption{Random vs. Halton sampling for 100 test iterations.
            The blue area is the largest rectangle without a test.
        X axis is the walking speed of the pedestrian ([0.5, 10] m/s).
        Y axis is the distance from the car when the pedestrian starts crossing ([5, 60] m).
        Failing tests (collisions) are marked with a red cross, passing tests are labelled with a green dot.
    }\label{fig:PedestrianHalton}
\end{figure}

\begin{table}
\centering
\caption{Random vs. Halton sampling for the pedestrian crossing experiment over various test iterations. The test parameters are the walking speed of the pedestrian ([0.5, 10] m/s) and distance from the car when the pedestrian starts crossing ([5, 60] m).}
\label{tbl:crossing}
\begin{tabular}{rrrrr}
\toprule
         & \multicolumn{2}{c}{Dispersion values} & \multicolumn{2}{c}{Failing tests}\\
\cmidrule(r){2-3}\cmidrule(r){4-5}
\# tests & Random   & Halton  & Random  & Halton    \\ \midrule
 50      & $0.200$  & $0.083$ & $6\%$     & $8\%$       \\
100      & $0.105$  & $0.041$ & $6\%$     & $8\%$       \\
200      & $0.051$  & $0.029$ & $7\%$    & $8\%$      \\
400      & $0.025$  & $0.011$ & $8.75\%$    & $8.25\%$      \\ \bottomrule
\end{tabular}
\end{table}

\paragraph{Dynamic Pedestrian Behavior (low dispersion sequences revisited).}
This case study has already been presented in the main paper.
We now present results that underline the importance of low-dispersion sequences and how coverage improves with more test iterations.
Note that the parameter ranges here are different (wider) than the study presented in the main paper. However, the SUT is the same.
Figure~\ref{fig:PedestrianHalton} demonstrates the advantage of low-dispersion sampling over random sampling. 
Samples are more spread out for the Halton sequence (low-dispersion).
In Table~\ref{tbl:crossing}, we report the difference between random and Halton sampling for various numbers of test iterations.
Halton sampling gives much better dispersion and even leads to more failure cases being revealed (especially for fewer test iterations).

\begin{figure}
\centering
  	\includegraphics[width=0.75\textwidth]{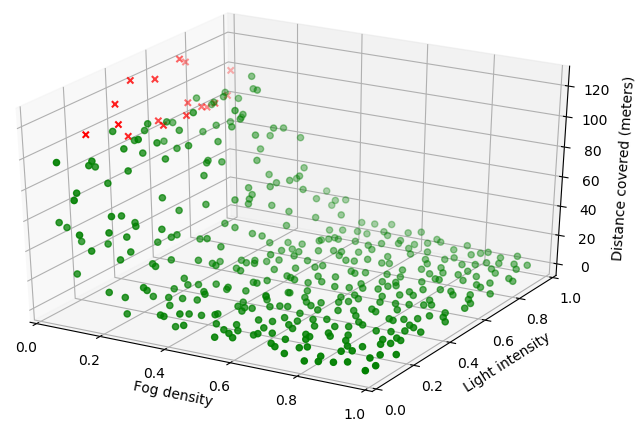}
  	\caption{Distance covered (Z-axis, [0,120] m) in changing fog (X-axis, $[0,1]$) and light (Y-axis, $[0,1]$) conditions tested with 400 iterations of the Halton sequence. 
Green dots and red crosses denote the absence or presence of a collision.
The car is trained with a fog density of 0 and light intensity of 0.5.}
  \label{fig:FogDensityvsDistanceHalton}
\end{figure}

\paragraph{Changing Environmental Settings.}
As mentioned in the main paper, reactive variables can be used to parameterize 
environment settings so as to describe a large class of configurations.
To demonstrate this, we train a model at fixed light intensity and no fog.
This case study is similar to the Adaptive Cruise Control case study presented in the main paper. 
Here, we analyse the autonomous vehicle's performance when the light intensity and fog density are varied.
We report the overall distance covered, and whether a collision happened.
Each test lasts 15 seconds.
Parameter values are generated using the Halton sequence.
Results are aggregated in Figure \ref{fig:FogDensityvsDistanceHalton}.
The car performs best around the parameter values it was trained on.
The distance that the car covers drops-off fairly quickly as fog density increases.
Perturbations to light intensity often lead to scenarios with collisions.

\begin{figure}
  \centering
    \begin{subfigure}{0.4\textwidth}
    \centering
        \includegraphics[width=0.7\textwidth]{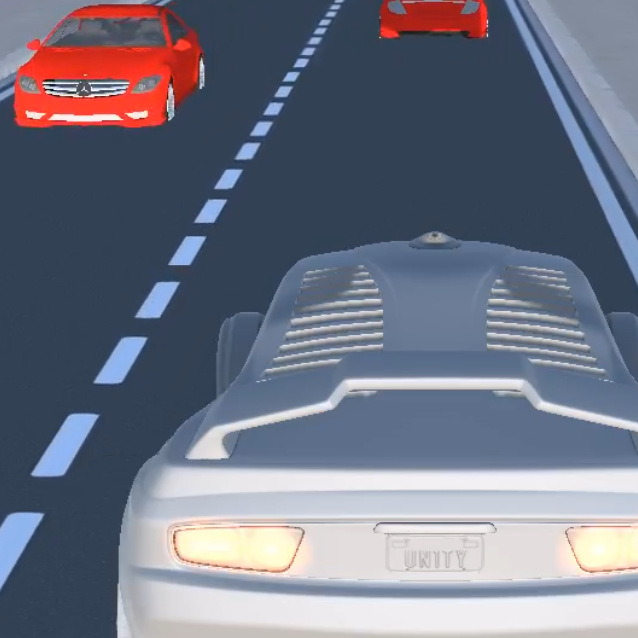}
        \caption{Test vehicle braking on seeing a red car coming from the opposite direction, even though there is a large distance to the lead car (car on the same lane).}
        \label{fig:brake_opp_car}
    \end{subfigure}
    \hfill
    \begin{subfigure}{0.55\textwidth}
    \centering
	\includegraphics[width=0.95\textwidth]{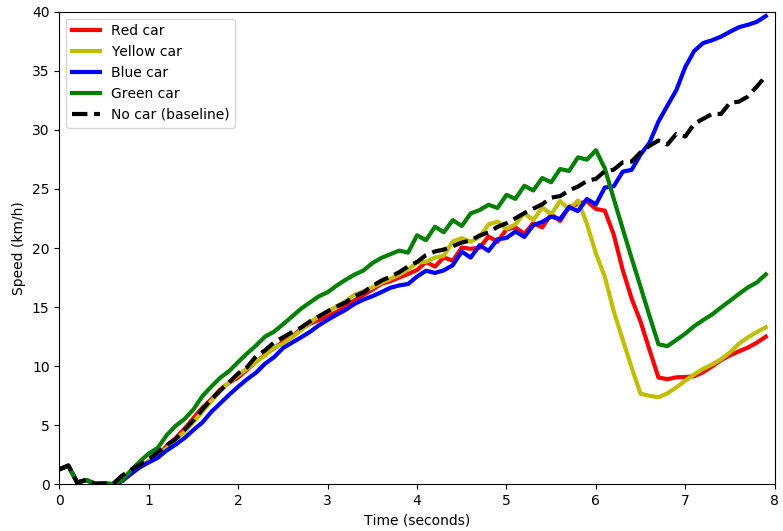}
  	\caption{Speed (X-axis, $[0,40]$ km/h) over time (Y-axis, $[0,8]$ sec) of car trained to follow a red car in the presence of another car coming from the opposite direction.
  	Depending on the color of the incoming car, the speed of the car changes vis-\`a-vis the baseline driving with no other car.
  	}
  	\label{fig:AverageSpeedTicks}
    \end{subfigure}
    \caption{Effect of features of geometric components.}
    \label{fig:features_geometric}
\end{figure}

\paragraph{Features of Geometric Components.}

For the case study above, the SUT is trained to follow a red lead car driving in front of it on a two-lane road.
Under ideal conditions (conditions under which the SUT is trained), it is observed that the autonomous vehicle indeed follows the red lead car while maintaining a safe distance and not colliding with it.
We now test how the SUT reacts to cars coming from the other direction. 
Though the test vehicle's throttle should not be affected by cars coming from the other direction, it is likely that the SUT learnt to slow the car down when there are several red pixels in the camera image.
Indeed, this seems to be the case. 
When we test with a red car coming from the other direction, our autonomous vehicle slows down in response to this car being close (see Figure~\ref{fig:brake_opp_car}). 
Speed is picked up again once this car passes. 
Perhaps more surprisingly, the vehicle also slows down when the car coming from the other direction is yellow or green, but has no affect when the car is blue. 
Figure~\ref{fig:AverageSpeedTicks} has plots of speed vs. time for the various cases, with the baseline being no car coming from the other direction.  

%

\section{Connections of road segments (and \opendrive output)}
\label{sec:conn_road_segments}

\begin{figure}
    \centering
    \begin{subfigure}{0.45\textwidth}
    \centering
        \includegraphics[width=0.8\textwidth]{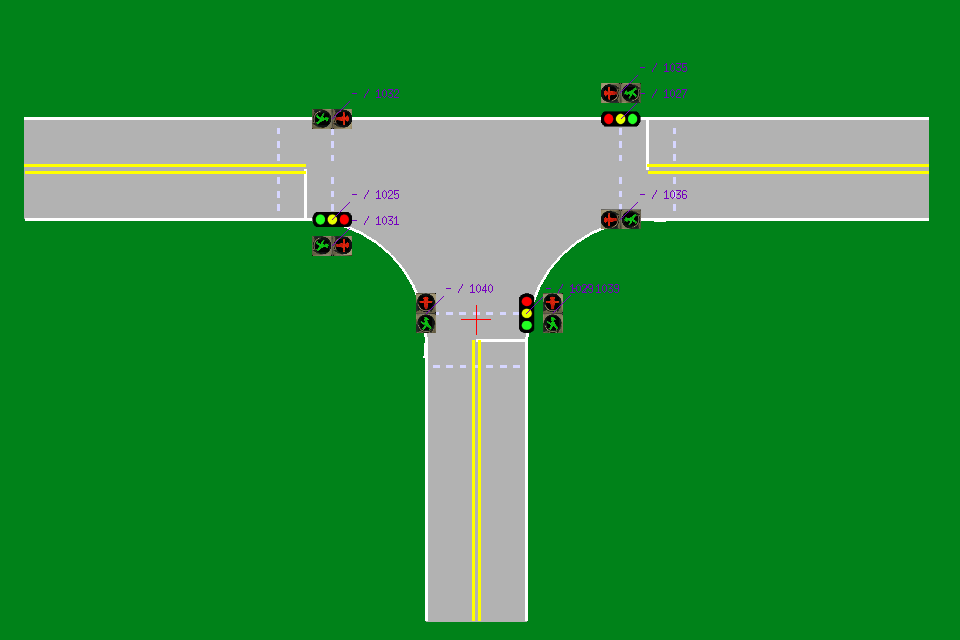}
        \caption{T-Intersection with 2 lanes, long road segments, and traffic and pedestrian lights.}
        \label{fig:tintersection-opendrive}
    \end{subfigure}
    \begin{subfigure}{0.45\textwidth}
    \centering
        \includegraphics[width=0.8\textwidth]{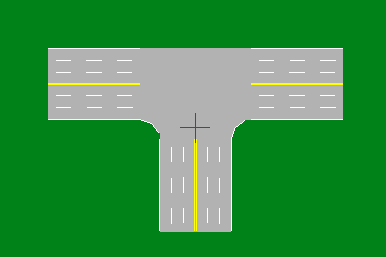}
        \caption{T-Intersection with 4 lanes, short road segments, and no traffic or pedestrian lights.}
        \label{fig:tintersection-opendrive_simple}
    \end{subfigure}
    \caption{Parameterized road segments outputted to \opendrive. Options to create vehicular and pedestrian traffic lights can also be arguments to the \lstinline{TIntersection} interface.
    }\label{fig:tintersections}
\end{figure}

In this section we provide a code example to demonstrate composition of 
road elements using the \lstinline{connect} operation between road elements.
\tool supports complex road elements such as cross-intersections, T-intersections, and roundabouts.
Connections can be established using the \lstinline{connect} method, that takes physical connection identifiers and road elements as arguments.
The connections are directed in order to compute the positions of the elements.
One road element becomes the parent and it's children are positioned relative to its position and the specified connection points.
After an object is connected, a new \emph{composite} road element which encapsulates all road elements along with requisite transformations (rotations and translations) is returned.
The following example shows how road segments can be connected into a road network.
\begin{lstlisting}
len = VarInterval(5, 100)
nlanes = VarInterval({2, 4})
// Create a parameterized T-intersection and three straight road elements (east, south, west)
t = TIntersection(nlanes:nlanes)
e = StraightRoadSegment(len:len, nlanes:nlanes)
s = StraightRoadSegment(len:len, nlanes:nlanes)
w = StraightRoadSegment(len:len, nlanes:nlanes)
// connect and get new composite object
net = t.connect((t.ONE, e, e.TWO),
                (t.TWO, s, s.ONE), 
                (t.THREE, w, w.ONE))
\end{lstlisting}

In this example, the T-intersection is not given a specific position or orientation.
It is therefore instantiated at the origin.
Road elements connecting to it are then positioned with respect to it.
After connection, the composite road element \lstinline{net} can be used for tests in simulation or to a standardized format (\opendrive).
Figure~\ref{fig:tintersections} shows some samples in the \opendrive viewer.

Connecting elements has two purposes.
First, it allows \tool to perform sanity checks like proper positioning of road elements.
Second, it creates an overlay graph of the road networks which can easily be followed by environment controlled vehicles.
When a road network is created, the runtime system of \tool checks that compositions of road elements and intersections are topologically and geometrically valid.
All road elements must be connected to a matching road correctly (for example, a 2-lane road segment cannot be connected to a 6-lane road segment directly),
there can be no spatial overlaps between road segments, and the positions of the connection points must match.

\begin{figure}
  \centering
  	\includegraphics[width=0.55\textwidth]{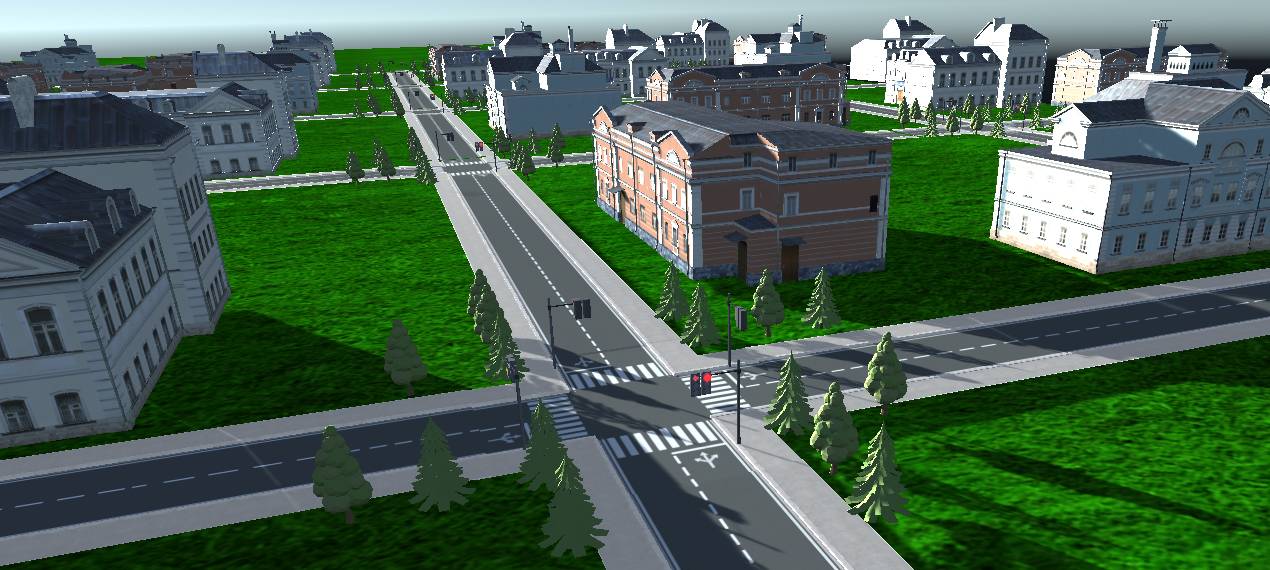}
  	\caption{A large grid world with several connected road elements viewed in the default 3D simulator.}
  \label{fig:GridWorld}
\end{figure}

In general, \tool inherits all programming features of the underlying imperative programming model as
well as reactive programming with streams.
Thus, one can build complex urban settings through composition and iteration.
For instance, the grid world shown in Figure~\ref{fig:GridWorld} was created by iterating a simple road network.

\vfill

{\small\medskip\noindent{\bf Open Access} This chapter is licensed under the terms of the Creative Commons\break Attribution 4.0 International License (\url{http://creativecommons.org/licenses/by/4.0/}), which permits use, sharing, adaptation, distribution and reproduction in any medium or format, as long as you give appropriate credit to the original author(s) and the source, provide a link to the Creative Commons license and indicate if changes were made.}

{\small \spaceskip .28em plus .1em minus .1em The images or other third party material in this chapter are included in the chapter's Creative Commons license, unless indicated otherwise in a credit line to the material.~If material is not included in the chapter's Creative Commons license and your intended\break use is not permitted by statutory regulation or exceeds the permitted use, you will need to obtain permission directly from the copyright holder.}

\medskip\noindent\includegraphics{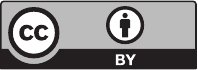}

\end{document}


\title{Supplementary Material for ``\tool: A Tool for Testing Autonomous Driving Systems''} 

\maketitle

\appendix
\section{Details of the Object Segmentation Experiment}
\label{app:objseg}

In this experiment, we test the detection rate of various road and
vehicle detection systems subjected to change several parameters
influencing the placement of the camera, traffic and environmental
effects. 
The experiment is designed to quickly traverse the search space, while
ensuring a low dispersion rate. The search space consists of both,
continuous and discrete parameters.
An experiment consists of 100 iterations, each described by a combination
of the following parameters. The values in parentheses give the interval
(continuous parameter) or the set of possible values (discrete parameter).

\begin{itemize}
\item Number of lanes (discrete, $\{2, 4\}$)
\item Number of cars (discrete, $\{2,3,..,9\}$)
\item Time of day (discrete, $\{Morning (0), Noon (1), Evening (2)\}$)
\end{itemize}

\begin{itemize}
\item Height of camera (continuous, $[1.9, 2.2]$)
\item Pitch of camera (continuous, $[-12, -10]$)
\item Focal length of camera (continuous, $[18, 22]$)
\end{itemize}

In each iteration, the scene (i.e., position of the ego car, road network, 
Zoning) stays the same, while the above mentioned parameters are Halton 
sampled. A hardcopy of what a camera would see is then captured and subjected 
to various tests.

These tests consist of road detection systems and vehicle detection system.
Each system operates on the same images, in the same order. For the vehicle 
detection, the baseline consists of all cars visible in the scene (and, in 
theory, visible to the ego car). Between iterations, the number may vary, but
the positions remain the same. We record the number of detected cars. 

For the road detection, the baseline consists of the road visible in the scene.
We record the true positive and the false positive percentage of road detected,
e.g. detected road which is indeed road and detected road which is something else,
e.g. sidewalks.

\tool is able to generate the baseline for both, vehicle detection and road 
detection, automatically.

We use the following implementations for the vehicle detection:
\begin{description}
    \item[SSD] Single Shot Multi-Box Detector \cite{vehicleDetectionSSD2015}, Code taken from \url{https://github.com/ndrplz/self-driving-car/tree/master/project_5_vehicle_detection}
    \item[Multinet] Multinet neural network, \cite{TeichmannWZCU16}, Code from \url{https://github.com/MarvinTeichmann/KittiSeg}
\end{description}

We use the following implementations for the road detection:
\begin{description}
    \item[SZ] Fully convolutional Network by \cite{Simonyan14verydeep}, code taken from  \url{https://github.com/ndrplz/self-driving-car/tree/master/project_12_road_segmentation}
    \item[LSD] Fully convolutional network for Semantic Segmentation \cite{LongSD14}, code taken from \url{https://github.com/JunshengFu/semantic_segmentation}
    \item[Multinet] Multinet neural network, \cite{TeichmannWZCU16}, Code from \url{https://github.com/MarvinTeichmann/KittiSeg}
\end{description}

For readability, only four decimal places are reported.

\begin{table*}
\caption{Car Detection}
\tiny
\begin{tabular}{lllllll|llll}
    Test &   \# Lanes & Cars     &                 Focal length    & Camera Height &   Camera Pitch&   Time of day    &   Detected Cars SSD  &   $[\%]$  &      Detected Cars Multinet & $[\%]$                      \\ \hline 
1    &             2 &           7  &                 18.8 &      2.05 & 10.6666  &         Morning       & 3 &    0.4285 &                    2&   0.2857         \\
2  &             4 &           8  &                     19.6 &     1.975 & 11.33333  &      Evening      & 2 &    0.2500 &                    3&   0.3750         \\
3  &             2 &           7  &                     20.4 &     2.125 & 10.22222  &      Morning      & 2 &    0.2857 &                    3&   0.4285         \\
4  &             4 &           8  &                     21.2 &    1.9375 & 10.88889  &      Evening      & 3 &    0.3750 &                    3&   0.3750         \\
5  &             4 &           5  &                    18.16 &    2.0875 & 11.55556  &      Noon      & 1 &    0.2000 &                    2&   0.4000         \\
6  &             2 &           7  &                    18.96 &    2.0125 & 10.44444  &      Noon      & 1 &    0.1428 &                    2&   0.2857         \\
7  &             4 &           3  &                    19.76 &    2.1625 & 11.11111  &      Evening      & 2 &    0.6666 &                    2&   0.6666         \\
\hline
\end{tabular}
\end{table*}

\begin{table*}
\caption{Road Detection}
\tiny
\begin{tabular}{lllllll|llllll}
Test  &  \# Lanes & Cars& Focal length & Camera Height & Camera Pitch & Time of day & SZ $[\%]$  &  False Positive SZ $[\%]$ & LSD $[\%]$ &    False Positive LSD $[\%]$ &  Multinet $[\%]$ & False Positive Multinet $[\%]$ \\ \hline 
 1    &             2 &          7  &        18.8000 &   2.0500 & 10.6666  &       Morning   &             0.4539&     0.0005 &             0.6329 &    0.0039  &              0.2532 &         0.0238 \\
 2 &             4 &          8  &           19.6000 &   1.9750 & 11.3333  &       Evening   &          0.3749&     0.0244 &             0.0397 &    0.0340  &              0.1400 &         0.0008 \\
 3 &             2 &          7  &           20.4000 &   2.1250 & 10.2222  &       Morning   &          0.7306&     0.0373 &             0.0269 &    0.0611  &              0.2155 &         0.0000 \\
 4 &             4 &          8  &           21.2000 &   1.9375&  10.8888 &        Evening   &           0.3707&     0.0292 &             0.0386 &    0.0066  &              0.1099 &         0.0000 \\
 5 &             4 &          5  &           18.1600 &   2.0875&  11.5555 &        Noon   &           0.4261&     0.0033 &             0.5243 &    0.0251  &              0.2826 &         0.0067 \\
 6 &             2 &          7  &           18.9600 &   2.0125&  10.4444 &        Noon   &           0.5016&     0.0926 &             0.0188 &    0.0705  &              0.1318 &         0.0000 \\
 7 &             4 &          3  &           19.7600 &   2.1625&  11.1111  &       Evening   &          0.4130&     0.0051 &             0.0570 &    0.0076  &              0.1906 &         0.0000\\
\hline
\end{tabular}
\end{table*}

\bibliography{../Bibliography}


\begin{adjustbox}{angle=90}
\tiny
\begin{tabular}{lllllllllllllllll}
Iteration &    NumberofLanes & VisibleCars& DetectedCarsSSD  &   PercentageSSD   &      DetectedCarsKittyBox & PercentageKittyBox & PercentageDetectedRoad RS  &  FalsePositiveRoad RS & PercentageDetected Road SS &    FalsePositiveRoad SS&  PercentageDetected Road KittySeg& FalsePositiveRoad KittySeg &Focal    & Height &   Pitch&   Light  \\ \hline 
0 &             2 &          7  &                 3 &    0.428571428571429 &                    2&   0.285714285714286 &           0.453999142734676&     0.000514470878574 &             0.632936133733391 &    0.003903917401327  &              0.253240462923275 &         0.023834443081783 &     18.8 &      2.05 & 10.66667  &       0 \\
1 &             4 &          8  &                 2 &                 0.25 &                    3&               0.375 &           0.374965255730573&     0.024481609562398 &             0.039701543538076 &    0.034077372816631  &              0.140086344290847 &         0.000834994209713 &     19.6 &     1.975 & 11.33333  &       2 \\
2 &             2 &          7  &                 2 &    0.285714285714286 &                    3&   0.428571428571429 &           0.730676016881992&     0.037347755041528 &             0.026912585930445 &    0.061157316440917  &              0.215572531745885 &                  0.00E+00 &     20.4 &     2.125 & 10.22222  &       0 \\
3 &             4 &          8  &                 3 &                0.375 &                    3&               0.375 &           0.370728622504595&     0.029211640438612 &             0.038605107813296 &    0.006600251543411  &              0.109926992571303 &                         0 &     21.2 &    1.9375 & 10.88889  &       2 \\
\end{tabular}
\end{adjustbox}